\documentclass[12pt,urlcolor=black,linkcolor=black]{article} 
\pdfoutput=1
\usepackage{cite}
\usepackage{amsmath, amsthm, amssymb,slashed, url}
\usepackage{pifont}

\usepackage{ifpdf}
\ifpdf
  \usepackage[pdftex]{graphicx}
  \usepackage{epstopdf}
\else  
  \usepackage[dvips]{graphicx}
\fi

\usepackage{bbm}
\def\q{{\mathbbm{q}}}
\def\lk{{\ell k}}
\def\vert{\mathrm{Vert}}

\usepackage[usenames,dvipsnames]{color}
\usepackage[colorlinks,citecolor=black]{hyperref}

\parskip=\baselineskip

\def\bar{\overline}
\def\tilde{\widetilde}
\def\hat{\widehat}

\numberwithin{equation}{section}

\newcommand{\Z}{{\mathbb Z}}
\newcommand{\R}{{\mathbb R}}
\newcommand{\C}{{\mathbb C}}

\newcommand{\Q}{{\mathbb Q}}

\def\CH{{\mathcal H}}
\def\CI{{\mathcal I}}

\def\CL{{\mathcal L}}
\def\CM{{\mathcal M}}
\def\CN{{\mathcal N}}

\newcommand{\bea}{\begin{eqnarray}}
\newcommand{\eea}{\end{eqnarray}}
\newcommand{\be}{\begin{equation}}
\newcommand{\ee}{\end{equation}}

\definecolor{ao}{rgb}{0.13, 0.55, 0.13}

\newcommand{\cmark}{{\color{ao}\ding{51}}} 

\begin{document}

\def\aa{{\sf a}}
\def\bb{{\sf b}}
\def\tr{{\mathrm{tr}}}
\def\Tr{{\rm Tr \,}}
\def\diag{{\mathrm{diag}}}
\def\Re{{\mathrm{Re}}}
\def\Im{{\mathrm{Im}}}

\begin{titlepage}
\begin{flushright}
\end{flushright}
\vskip 1.5in
\begin{center}
{\bf\Large{Cobordism invariants from BPS $q$-series}}
\vskip
0.5cm {Sergei Gukov,$^{1,2}$ Sunghyuk Park,$^2$ Pavel Putrov$^3$} \\
\vskip.8cm
{\small{ \textit{$^1$ Walter Burke Institute for Theoretical Physics, California Institute of Technology,}\vskip -.4cm
{\textit{Pasadena, CA 91125, USA}}} \vskip-.4cm 
\small{ \textit{$^2$ Division of Physics, Mathematics and Astronomy, California Institute of Technology,}\vskip -.4cm
{\textit{Pasadena, CA 91125, USA}}} \vskip-.4cm 
{\small{ \textit{$^3$ {International Centre for Theoretical Physics,}\vskip -.4cm
{\textit{Strada Costiera 11, Trieste 34151 Italy}}}
}}}
\end{center}
\vskip 0.5in
\baselineskip 16pt
\abstract{Many BPS partition functions depend on a choice of additional structure: fluxes, Spin or Spin$^c$ structures, {\it etc.} In a context where the BPS generating series depends on a choice of Spin$^c$ structure we show how different limits with respect to the expansion variable $q$ and different ways of summing over Spin$^c$ structures produce different invariants of homology cobordisms out of the BPS $q$-series.}
\date{August, 2020}
\end{titlepage}

\section{Introduction}

In supersymmetric theories, BPS states --- named after Bogomol'nyi, Prasad, and Sommerfield --- often have a geometric interpretation as volume minimizing submanifolds or solutions to partial differential equations. This is especially common among theories that can be realized in string theory, and string dualities typically offer several equivalent perspectives on BPS objects.

Counting such minimal or extremal objects then leads to a generating series that captures a lot of interesting information about the underlying geometry.
One well known example of such counting problem involves curve counting in Calabi-Yau manifolds. It can be formulated in such a way that, in each topological class, the result is an integer \cite{MR1634503,Gopakumar:1998jq,Aganagic:2003db,Iqbal:2003ds,Maulik:2003rzb} and altogether these integer invariants can be conveniently packaged into a generating $q$-series.
Another prominent example involves counting solutions to the Vafa-Witten equations on 4-manifolds \cite{Vafa:1994tf}. When realized in string theory, these two seemingly different geometric incarnations of BPS states can be related (see {\it e.g.} \cite{Minahan:1998vr,Gukov:2016zqu,Gholampour:2017bxh,MR3882207} and references therein).

Sometimes, the underlying geometric setup requires a choice of a certain additional structure, and the resulting generating series depends on that choice.
For example, in the case of Vafa-Witten gauge theory, one finds a collection of $q$-series invariants of 4-manifolds
\be
Z_{VW}^{(v)} (M_4,q) \; = \; q^{h_v} \left( a_0^{(v)} + a_1^{(v)} q + a_2^{(v)} q^2 + \ldots \right)
\ee
labeled by 't Hooft fluxes $v$.
When $b_2^+ (M_4) > 1$, this collection transforms as a vector-valued modular form under the modular group $SL(2,\Z)$ or its congruence subgroup.

The analogue of Vafa-Witten invariants for 3-manifolds, denoted $\hat Z_b (M_3,q)$, also depends on the additional choice --- for gauge group $G=SU(2)$, the choice of Spin$^c$ structure on $M_3$ --- and also can be interpreted as counting BPS states \cite{Gukov:2016gkn}, either in terms of curve counting, or as the half-index of 3d $\CN=2$ theory with 2d $(0,2)$ boundary conditions \cite{Gadde:2013wq}, or as counting solutions to Kapustin-Witten equations \cite{MR2852941}.
(See \cite{Gukov:2017kmk} for a comprehensive survey of all these perspectives and their interrelations.)
These $q$-series invariants
\be
\widehat Z_b (M_3,q) \; = \; q^{\Delta_b} \left( a_0^{(b)} + a_1^{(b)} q + a_2^{(b)} q^2 + \ldots \right) \quad \in \; q^{\Delta_b} \Z [[q]]
\label{Zhatqseries}
\ee
labeled by $b \in \text{Spin}^c (M_3)$ are relatively easy to compute, which makes the study of their structure very accessible.\footnote{The fact that $\hat Z_b (M_3,q)$ is labeled by Spin$^c$ structures was carefully explained in \cite{Gukov:2019mnk} and recently justified further from a different perspective \cite{Gukov:2020lqm}, by realizing $\hat Z_b (M_3,q)$ as a Rozansky-Witten theory with the target space based on the affine Grassmannian.} It is natural to ask, then, how much topological information about $M_3$ is contained in $\hat Z_b (M_3,q)$, in particular in the dependence on $b \in \text{Spin}^c (M_3)$.

Topological invariants of a seemingly different origin recently played an important role in the study of quantum states of matter. Perhaps the most well known is the classification of fermionic symmetry protected topological (SPT) phases, given by the Spin cobordism groups \cite{Kapustin:2014dxa,Freed:2016rqq}:
\be
\begin{array}{c|c|c|c|c|c|c|c|c|c|c|c}
	d & 0 & 1 & 2 & 3 & 4 & 5 & 6 & 7 & 8 & 9 & \ldots \\
	\hline
	\phantom{\Big(}\Omega_{d}^{\text{Spin}} \phantom{\Big)} & ~\Z~ & ~\Z_2~ & ~\Z_2~ & ~0~ & ~\Z~ & ~0~ & ~0~ & ~0~ & ~\Z^2~ & ~\Z_2^2~ & ~\ldots~
\end{array}
\ee
For example, the group in dimension $d=2$ classifies two inequivalent SPT phases that can be represented by Kitaev's spin chain with only fermion number parity symmetry.
Note, the group in dimension $d=3$, relevant to 3-manifolds, is trivial. This does not mean, however, that the story ends here.
It means that every 3-manifold $M_3$ equipped with a Spin structure $s \in \text{Spin} (M_3)$ can be realized as a boundary of a Spin 4-manifold $M_4$ (with a Spin structure that extends $s$). Since by a theorem of Rokhlin the signature of a closed Spin 4-manifold is divisible by 16, this means that the mod 16 reduction of $\sigma (M_4)$ is independent of $M_4$ and
\be
\mu (M_3, s) \; := \; \sigma (M_4) 
\qquad \text{mod}~16
\label{Rokhlindef}
\ee
is a topological invariant of $(M_3,s)$ called the Rokhlin invariant.\footnote{Another choice of conventions used in the literature is $\mu (M_3, s) := \frac{\sigma (M_4)}{8}$ mod 2.}

Another way to make cobordisms in $d=3$ non-trivial is to restrict to integral homology spheres, $H_* (M_3;\Z) \cong H_* (S^3; \Z)$, and, similarly, consider cobordisms with $H_* (M_4;\Z) \cong H_* (S^3 \times [0,1]; \Z)$. The equivalence classes of integral homology spheres with respect to such cobordisms define an abelian group, called the homology cobordism group $\Theta^3_{\Z}$, which in some sense is a complete opposite of $\Omega_{3}^{\text{Spin}}$: not only is $\Theta^3_{\Z}$ infinite, but it is infinitely generated. A natural way to study this complicated group is via invariants of homology cobordisms.
This is where the Rokhlin invariant \eqref{Rokhlindef} makes its another appearance. Since integral homology spheres have unique Spin structure, there is no dependence on Spin structure and $\frac{\mu}{8}$ provides a homomorphism $\Theta^3_{\Z} \to \Z / 2\Z$.
For example, the Poincar\'e sphere $P = \Sigma (2,3,5)$ can be realized as a boundary of a negative definite 4-manifold with intersection form $-E_8$. Therefore, according to \eqref{Rokhlindef}, the Rokhlin invariant is $\mu (P) = 8$ mod 16 and it follows that the Poincar\'e sphere does not bound a homology ball.
Other invariants of homology cobordisms include:
\begin{itemize}

\item Fr{\o}yshov invariant $h : \Theta^3_{\Z} \to \Z$ (can be used to show that a connected sum of any number of $P$ is non-trivial in $\Theta^3_{\Z}$) \cite{MR1910040};

\item correction terms $d(M_3)\in \Q$ of Ozsv\'{a}th and Szab\'{o} defined via Heegaard Floer homology \cite{MR1957829};

\item invariants $\alpha (M_3)$, $\beta (M_3)$, $\gamma (M_3)$ introduced by Manolescu \cite{MR3402697}.

\end{itemize}

Our main interest here is to explore the relation between these delicate invariants and the BPS state counting. In particular, we shall focus on the Rokhlin invariant and the correction terms $d (M_3)$. Just as the Rokhlin invariant, $d (M_3)$ can be defined for an arbitrary 3-manifold, except that a choice of Spin$^c$ structure is required. There are certain similarities between $\mu (M_3,s)$ and $d (M_3,b)$, with $b \in \text{Spin}^c (M_3)$. For example, both change signs under the orientation reversal,
\begin{subequations}
\label{orientation}
\be
\mu (- M_3,s) \; = \; - \mu (M_3,s)
\ee
\be
d (- M_3,b) \; = \; - d (M_3,b)
\ee
\end{subequations}
and in what follows we will use this property a number of times. In comparison, the behavior of the BPS $q$-series \eqref{Zhatqseries} under the orientation reversal is very non-trivial in general \cite{Cheng:2018vpl}. However, the leading $q$-powers, $\Delta_b (M_3) \in \mathbb{Q}$, transform in a rather simple way, similar to \eqref{orientation}:
\be
\Delta_b (- M_3) \; = \; - \Delta_b (M_3).
\label{Deltaorientation}
\ee
Therefore, it is natural to ask if there is a relation between these invariants or, more generally, between $\hat Z_b (M_3,q)$ and $\mu (M_3,s)$ or $d (M_3, b)$. Answering this question could help to understand better the topological information contained in the BPS counting invariants, analogous to the way BPS invariants encode information about flat connections on $M_3$ \cite{Gopakumar:1997dv}. And, on the other hand, it could benefit topology by providing new bridges and connections between invariants like $\mu (M_3,s)$ and $d (M_3, b)$, realized as different limits of the BPS $q$-series.

More pragmatically, the goal of this paper is to study the relation between the following invariants, realized as limiting values of the BPS $q$-series \eqref{Zhatqseries}:
\begin{itemize}
\item $\{ \Delta_b (M_3) \}$ is a list of rational numbers labeled by $\text{Spin}^c$ structures on $M_3$,
	
\item $\{ d (M_3, b) \}$ is a list of rational numbers also labeled by $b \in \text{Spin}^c (M_3)$, and
	
\item $\{ \mu (M_3, s) \}$ is a list of $\Z / 16 \Z$ valued integers labeled by $\text{Spin}$ structures on $M_3$.
\end{itemize}
We can not stress enough that signs, factors of 2, and orientation conventions are absolutely crucial for this study. Since such details are sometimes ignored or omitted for the sake of simplifying the exposition, part of our motivation here is to carefully restore all such factors.
In particular, it was noted \cite{Gukov:2016gkn} that $\Delta_b (M_3)$ appear to be related to $d (M_3, b)$, with the same value of $b \in \text{Spin}^c (M_3)$. Part of our motivation here is to clarify this relation and to make it precise. We find\footnote{Note that modulo $2$ the correction terms of \cite{MR1957829} reduce to a more classical invariant (considered {\it e.g.} in \cite{APS-II}):
$$
d(M_3,b)\; = \; \frac{c_1(\tilde{b})^2-\sigma(M_4)}{4} \mod 2
$$
where $M_4$ is a 4-manifold with boundary $\partial M_4=M_3$ and a Spin$^c$ structure $\tilde{b}$ that extends $b = \tilde{b}|_{M_3}$.}
\be
\Delta_b (M_3) \; = \; \frac{1}{2} - d (M_3,b)
\qquad \text{mod}~1
\label{dvsDeltamain}
\ee
and
\be
\exp \left( - 2 \pi i \frac{3 \mu (M_3,s)}{16} \right)
\; = \; \sum_{b}
c^{\text{Rokhlin}}_{s,b} \; \hat Z_b (M_3, q) \Big|_{q=i}
\label{muvsZmain}
\ee
with simple ``universal'' coefficients $c^{\text{Rokhlin}}_{s,b}$ determined by the linking form on $H_1 (M_3;\Z)$, whose explicit form will be given below.

The rest of the paper is organized as follows.
In section \ref{sec:Spinsums} we provide further motivation for \eqref{muvsZmain} and discuss how various 3-manifold invariants can be constructed by playing with Spin$^c$ structure in \eqref{Zhatqseries}.
(In section \ref{sec:Spinsums} we will not discuss the relation between $\hat Z$-invariants and correction terms of $M_3$; a reader interested in this connection can skip directly to section~\ref{sec:surgeries}.)
Then, in section \ref{sec:surgeries}, we examine explicit examples of surgeries on knots which, in particular, help to determine the exact coefficients in \eqref{dvsDeltamain}--\eqref{muvsZmain}.
We conclude in section \ref{sec:conclusions} with some thoughts on possible applications and future directions.

\section{Different ways of combining Spin$^c$ structures}
\label{sec:Spinsums}

By taking various limits and combinations of the $q$-series invariants \eqref{Zhatqseries}, one can reproduce many other invariants of 3-manifolds, defined independently. This will be the main theme of this note.

For example, by summing over Spin$^c$ labels and taking $q$ to be a root of unity, one can obtain the familiar Witten-Reshetikhin-Turaev (WRT) invariants of $M_3$. On the other hand, by {\it not} summing over Spin$^c$ labels --- or, rather, replacing Spin$^c$ labels with their close cousins, as we explain shortly --- and taking the limit $q \to 1$ one can obtain the inverse Turaev torsion of $M_3$.
More precisely, the relation to the WRT invariants\footnote{We use the normalization of the WRT invariant such that $\mathrm{WRT}(S^3,k)=1$, which is standard in the mathematical literature. Note that the partition function of $SU(2)_k$ Chern-Simons theory, $Z_{SU(2)_k}$, is naturally normalized such that $Z_{SU(2)_k}(S^2\times S^1)=1$ instead, but has $Z_{SU(2)_k}(S^3)=\sqrt{\frac{2}{k}}\sin\frac{\pi}{k}$.} is usually written in the form\footnote{for simplicity, stated here for a 3-manifold with $b_1 (M_3) = 0$; generalization to $b_1 (M_3) > 0$ is not much more complicated \cite{Gukov:2017kmk,Chun:2019mal}, but will not be need here.}
\be
\text{WRT} ( M_3,k )
\; = \;
\lim_{q \to \exp \frac{2\pi i}{k}}
\;
\frac{1}{2(q^{\frac{1}{2}}-q^{-\frac{1}{2}})}
\sum_{b}
\underbrace{ \sum_a e^{2\pi i k \ell k (a,a)} S_{ab}}_{\text{Gauss sum}} \, \hat Z_b (q)
\label{WRTfromZhat}
\ee
with coefficients $S_{ab}$ expressed in a simple way through the linking form on $H_1 (M_3)$.
Thus, for manifolds with $H_1 (M_3) = \Z_p$, which will be the case for many of our examples below, the explicit form of these coefficients is \cite{Gukov:2016gkn}:
\be
S_{ab} \; = \;
\frac{1}{\sqrt{p}}
\left(- \frac{1+(-1)^{p+1}}{2} \delta_{a,0} + 2 \cos \frac{4\pi a b}{p} \right)
\ee
where $a,b = 0, \ldots, \lfloor \frac{p}{2} \rfloor $.
Note, this expression can be used for both even and odd values of $p$.
More precisely, this is the {\it folded} form, in which $\hat Z_b (M_3,q)$ are labeled by $b \in \text{Spin}^c (M_3) / \Z_2$.
While this form is more economical, sometimes it is more convenient to work with the {\it unfolded} version, where $\hat Z_b (M_3,q)$ are labeled by $b \in \text{Spin}^c (M_3)$ and the convolution with $S_{ab}$ simply implements the Fourier transform on $b \in \text{Spin}^c (M_3)$. For example, for $H_1 (M_3) = \Z_p$ the unfolded version of $S_{ab}$ looks much simpler
\be
S_{ab} \; = \; \frac{1}{\sqrt{p}} e^{4\pi i \frac{ab}{p}} \,, \qquad a,b = 0, \ldots, p-1.
\ee
It is this unfolded version that gives the inverse Turaev torsion if one of the labels is {\it not} summed over and is kept open \cite{Chun:2019mal}:
\be
\frac{1}{\text{Turaev torsion }(a)} \; = \;
\lim_{q \to 1}
\;
\sum_{b} S_{ab} \hat Z_b (q).
\label{inverseTuraev}
\ee

Whether a problem at hand favors the folded or unfolded version, the relation \eqref{WRTfromZhat} can be written in a more succinct form:
\be
\text{WRT} (M_3,k) \; = \; \sum_{b}
c^{\text{WRT}}_b \; \hat Z_b (q) \Big|_{q \to e^{\frac{2 \pi i}{k}}}.
\label{WRTviac}
\ee
Indeed, while the original form \eqref{WRTfromZhat} is more illuminating for comparison to Chern-Simons theory on $M_3$, for other applications it may be more convenient to sum over the label $a$ that makes no appearance either on the left-hand side or in $\hat Z_b (M_3,q)$. Moreover, as indicated in \eqref{WRTfromZhat}, the sum over $a$ is a Gaussian sum (or, a combination of Gaussian sums in the folded version) and performing this sum gives the coefficients $c^{\text{WRT}}_b$ that depend on $k$, $b$, and the linking form on $H_1 (M_3)$.

Then, written in the form \eqref{WRTviac}, this relation makes it clear that the coefficients $c^{\text{WRT}}_b$ play the role analogous to that of $S_{ab}$ in \eqref{inverseTuraev}. In both cases, the choice of these coefficients determines which topological invariant we get from the ``building blocks'' $\hat Z_b (M_3,q)$.
It is, therefore, natural to ask for other relations similar to \eqref{inverseTuraev} and \eqref{WRTviac}.

Another example of such relation involves a sum over $b \in \text{Spin}^c (M_3)$ that is quadratic rather than linear in $\hat Z_b (M_3,q)$,
\be
\CI_{3d} (M_3,q) \; = \; \sum_{b \in \text{Spin}^c (M_3)}
c^{\text{Index}}_{b} \; \hat Z_b (M_3,q) \, \hat Z_b (- M_3,q)
\label{3dindex}
\ee
with $c_b^{\text{Index}} = \frac{1}{2} | \text{Stab}_{\Z_2} (b) |$.
This way of combining BPS $q$-series gives the superconformal index of 3d $\CN=2$ theory obtained by compactifying 6d fivebrane theory on $M_3$ \cite{Gukov:2017kmk}.
Note, unlike \eqref{orientation} and \eqref{Deltaorientation}, the index \eqref{3dindex} is manifestly invariant under the orientation reversal.
Using modular properties of the BPS $q$-series \cite{Cheng:2018vpl} one can write the explicit form of the 3d index {\it e.g.} for $M_3 = \Sigma (2,3,7)$:
\be
\CI_{3d} (q) \; = \;
1-q^2+q^3-q^5-2 q^6+2 q^7-2 q^8-q^{10}+q^{11}-4 q^{12}+2 q^{13} +\ldots
\ee

One of the goals in this paper is to add to the list of relations \eqref{inverseTuraev}--\eqref{3dindex} another way to assemble $\hat Z_b (M_3,q)$ together that gives the Rokhlin invariant of $M_3$.
Since the latter depends on $s \in \text{Spin} (M_3)$, whereas $\hat Z_b (M_3,q)$ depends on $b \in \text{Spin}^c (M_3)$, the coefficients in such relation depend on both $s$ and $b$:
\be
\exp \left( - 2 \pi i \frac{3 \mu (M_3,s)}{16} \right)
\; = \; \sum_{b}
c^{\text{Rokhlin}}_{s,b} \; \hat Z_b (M_3, q) \Big|_{q=i}.
\label{muvsZ}
\ee
Using the behavior of $\hat Z_b (M_3, q)$ under the orientation reversal \cite{Cheng:2018vpl}, we can write this relation in the following equivalent form
\be
\exp \left( 2 \pi i \frac{3 \mu (M_3,s)}{16} \right)
\; = \; \sum_{b}
c^{\text{Rokhlin}}_{s,b} \; \hat Z_b (- M_3, q) \Big|_{q=i}
\label{muvsZorientation}
\ee
which can be also obtained from (\ref{orientation}a) and is therefore consistent with the behavior of Rokhlin invariant under the change of orientation.

Note, averaging the left-hand side of \eqref{muvsZ} over $s \in \text{Spin} (M_3)$ gives the WRT invariants of $M_3$ at $k=4$ \cite{MR1117149}, which was in part our motivation for exploring \eqref{muvsZ}. The coefficients in \eqref{muvsZ} therefore must satisfy the sum rules $\sum_{s} c^{\text{Rokhlin}}_{s,b} = c^{\text{WRT}}_{b,k=4}$, for any $b \in \text{Spin}^c (M_3)$.
Consequently, when $M_3$ is a $\Z_2$ homology sphere, {\it i.e.} admits a unique Spin structure $s=0$, we can simply take $c^{\text{Rokhlin}}_{0,b} =
{c^{\text{WRT}}_{b,k=4}}$.
Properties like this and specific examples ({\it cf.} the next section) can quickly determine the coefficients $c^{\text{Rokhlin}}_{s,b}$ in \eqref{muvsZ}.
For example, in the simplest case of integral homology spheres we have
\begin{subequations}
\label{cRokhlin}
\be
H_1 (M_3) \; = \; 0: \qquad\qquad
c^{\text{Rokhlin}} \; = \; \frac{1}{i \sqrt{8}}.
\ee
More generally, for $H_1 (M_3) = \Z_p$ with $p$ odd,
\be
H_1 (M_3) \; = \; \Z_p \quad (p~\text{odd}): \qquad\qquad
c^{\text{Rokhlin}}_b \; = \;
\frac{1}{8}
\sum_{n=0}^{7}
e^{- \pi i \frac{(np - 2b)^2 + 2p}{8p}}
\ee
where $b = 0, \ldots, p-1$.
The case of $H_1 (M_3) = \Z_p$ with $p$ even is the first instance where new examples are used to determine
\be
H_1 (M_3) \; = \; \Z_p \quad (p~\text{even}): \qquad
c^{\text{Rokhlin}}_{s,b} \; = \; \frac{1}{4}
e^{- \pi i \frac{p+2(b+sp/2)^2}{4p} }
\left( 1 + (-1)^{b} e^{(s-1) \frac{\pi i}{2} p} \right)
\ee
with $s=0,1$ and $b = 0, \ldots, p-1$.
And, for general $H_1 (M_3)$ with $b_1 (M_3) = 0$, one can use examples of plumbed manifolds considered in section \ref{sec:spin-WRT-plumbed} to determine
\be
b_1 (M_3) \; = \; 0: \qquad\qquad
c^{\text{Rokhlin}}_{s,\sigma(s,b)} \; = \; \frac{1}{i \sqrt{8 |H_1 (M_3)|}} \sum_{a \in H_1 (M_3)} e^{- 2\pi i \,\lk (a,a) - 2\pi i\, \lk (a , b)}
\label{cRokhlin-general}
\ee
\end{subequations}
where $b \in H_1(M_3)$ and $\sigma(s,b)\in \text{Spin}^c(M_3)$ is the value of a canonical map $\sigma:\text{Spin}(M_3)\times H_1(M_3,\Z)\longrightarrow \text{Spin}^c(M_3)$ explained in more detail in \eqref{spinspincmap}.
For a fixed Spin structure $s$, this map provides a bijection between $\text{Spin}^c(M_3)$ and $H_1(M_3,\Z)$ and, therefore, can be inverted.

Note, the left-hand side of \eqref{muvsZ} is a special instance of the Spin refinement of the WRT theory, at level $k=4$ \cite{MR2371384,MR2457479}.
Therefore, eq. \eqref{muvsZ} itself can be understood as a relation of the type \eqref{inverseTuraev}--\eqref{3dindex} between Spin TQFT of \cite{MR2371384,MR2457479} and BPS $q$-series invariants.

Before we proceed to calculations, let us make a few general remarks about Spin and Spin$^c$ structures on 3-manifolds.
Recall, that $\text{Spin} (M_3)$ (resp. $\text{Spin}^c (M_3)$) is a torsor over $H^1 (M_3;\Z_2)$ (resp. $H^2(M_3,\Z)\cong H_1 (M_3; \Z)$), {\it i.e.} the difference between two Spin or Spin$^c$ structures is measured by elements of $H^1 (M_3;\Z_2)$ or $H^2(M_3,\Z)\cong H_1 (M_3; \Z)$, respectively.
We can use this to describe the Spin structure dependence in \eqref{Rokhlindef} and its behavior under surgery.
Indeed, suppose $M_3$ is obtained from $S^3$ by a surgery on a framed link $L \subset S^3$. This also defines a 4-manifold $M_4$, such that $\partial M_4 = M_3$, whose Kirby diagram is $L$. It consists of only 2-handles and a single 0-handle. Note, $M_4$ is oriented and simply-connected, but may not be Spin. Still, we can write an analogue of \eqref{Rokhlindef} by using a one-to-one correspondence between Spin structures on $M_3$ and characteristic surfaces $\Sigma$ inside $M_4$. Recall, that $\Sigma$ is called characteristic if $[\Sigma]$ is Poincar\'e dual to $w_2$, that is
\be
\Sigma \cdot x \; \equiv \; x \cdot x 
\qquad \text{mod}~2
\label{charcondition}
\ee
for every $x \in H_2 (M_4)$. For a 4-manifold represented by the Kirby diagram $L$, such characteristic surfaces can be conveniently identified with the characteristic sublinks of $L$. These are sublinks of $L$ which satisfy \eqref{charcondition} for every link component $x$ of $L$. Then, representing a Spin structure on $M_3$ by its characteristic surface (sublink), we can define the corresponding Rokhlin invariant \eqref{Rokhlindef} as
\be
\mu (M_3, s) \; = \; \sigma (M_4) - \Sigma \cdot \Sigma + 8 \, \text{Arf} (\Sigma)
\qquad \text{mod}~16
\label{muviacharsurf}
\ee
where $\text{Arf} (\Sigma)$ is the Arf invariant associated with the quadratic form on $H_1 (\Sigma; \Z_2)$, see {\it e.g.} \cite{MR1117149}. Note, the combination of the signature and the self-intersection of $\Sigma$ here is similar to the one in the index formula for a twisted Dirac operator on $M_4$. In the case of a smooth closed 4-manifold, this same combination appears as a Kervaire-Milnor obstruction for realizing a characteristic class by a smoothly embedded 2-sphere.

\subsection{Spin TQFT and plumbed 3-manifolds}
\label{sec:spin-WRT-plumbed}

\begin{figure}[tbp]
\centering 
\includegraphics[scale=2.5]{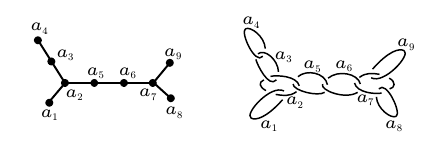}
\caption{An example of a plumbing graph $\Gamma$ and the corresponding framed link $\CL(\Gamma)$ in a three-sphere.}
\label{fig:plumbing}
\end{figure}

In this subsection we consider the Spin-refined version of the WRT invariant at level $k$. We focus on the case when a 3-manifold $M_3$ is realized by a plumbing, {\it i.e.} surgery on an arbitrary link of unknots, illustrated in Figure~\ref{fig:plumbing}. Using a generalized Gauss reciprocity formula we relate the Spin-refinement of the WRT invariant to the invariants $\hat{Z}_b$ in this case. Based on this, we then conjecture a relation between these two invariants for arbitrary rational homology spheres. While the connection to Spin TQFT is interesting in its own right, here it will help us to establish (\ref{cRokhlin-general}) in the relation between Rokhlin invariants and BPS $q$-series invariants.

Let us start by reviewing the definition of the Spin-refined version of the WRT invariant \cite{MR2457479,MR1117149}. As in the case of the ordinary WRT invariant, it is defined in terms of a Dehn surgery representation of a 3-manifold. Namely, let $M_3$ be a closed oriented 3-manifold obtained by a surgery on a framed link $\CL\subset S^3$. Denote by $L$ the number of link components of $\CL$ and by $Q$ its $L\times L$ linking matrix. The diagonal elements $Q_{II}=a_I\in \Z$ are the self-linking numbers of the individual components of $\CL$ labeled by $I$. In particular the numbers $a_I$  specify the framing of $\CL$. We will denote by $b_\pm$ the number of positive/negative eigenvalues of the linking matrix $Q$.

Let $J[\CL]_{n}\in \Z[q^{-1/2},q^{1/2}]$ be the colored Jones polynomial of $\CL$, where $n\in \Z_+^L$ is the color vector (so that $n_I\in \Z_+$ is the dimension of $SU(2)$ representation coloring the $I$-th component). It is normalized so that
\begin{equation}
    J[U_a]_n=q^{\frac{a(n^2-1)}{4}}\,\frac{q^\frac{n}{2}-q^{-\frac{n}{2}}}{q^\frac{1}{2}-q^{-\frac{1}{2}}}
\end{equation}
where $U_a$ is the unknot with framing $a\in \Z$.

To define the Spin-refined version of the WRT invariant it is useful to consider the following quantity associated to a framed link $\CL$ and a given ``mod 2 color'' $c\in \Z_2^L$:
\begin{equation}
F^{(c)}[\CL]:= \sum_{\scriptsize\begin{array}{c} 0 \leq n_I\leq 2k-1,\\ n_I=c_I\mod 2 \end{array}}' 
(J[\CL]_{n})
\Big|_{q=\q}\,
\prod_{I} \frac{\q^{n_I/2}-\q^{-n_I/2}}{\q^{1/2}-\q^{-1/2}},
\label{RT-sum}
\end{equation}
along with
\begin{equation}
F^\text{tot}[\CL]:= \sum_{ 0 \leq n_I\leq 2k-1}'
(J[\CL]_{n})
\Big|_{q=\q}\,
\prod_{I} \frac{\q^{n_I/2}-\q^{-n_I/2}}{\q^{1/2}-\q^{-1/2}},
\end{equation}
where the products are performed over all link components and the sums are performed over all their colors. The prime means that the ``singular'' terms $n_I=0,k$ should be omitted. Here and below $\q:=e^{\frac{2\pi i}{k}}$. In particular we have
\begin{equation}
    F^\text{tot}[U_{-1}]=\frac{2(2k)^{1/2}e^{-\frac{\pi i}{4}}\q^{3/4}}{\q^{1/2}-\q^{-1/2}},
\end{equation}
while $F^\text{tot}[U_{+1}]$ is its complex conjugate.
Finally, let $\varepsilon=(1,1,\ldots,1)\in \Z^L$ and define the refined $SU(2)$ WRT invariant as follows\footnote{We follow the normalization  convention of \cite{MR1117149}. The normalization of \cite{MR2457479} differs from the one in \cite{MR1117149} when $b_1(M_3)\neq  0$. This, however, does not affect the analysis done in this section, because we will assume that $M_3$ is a rational homology sphere.} \cite{MR2457479,MR1117149}:
\begin{equation}
\text{WRT}(M_3,k,s_c)=\frac{F^{(c+\varepsilon)}[\CL]}{F^\text{tot}[U_{+1}]^{b_+}F^\text{tot}[U_{-1}]^{b_-}}
\end{equation}
when $k$ is \textit{even}. One should distinguish two cases: 
\begin{itemize}

\item $k=0\mod 4$. Then $c\in \Z^L_2$ should satisfy $\sum_{J} Q_{IJ} c_J = Q_{II} \mod 2$ and the set of such solutions is in one-to-one correspondence with the set of Spin structures on $M_3$; $s_c\in \text{Spin}(M_3)$ is the Spin structure corresponding to $c$. This is the case we are mainly interested in.

\item $k=2\mod 4$. Then $c\in \Z^L$ should satisfy $\sum_{J} Q_{IJ}c_J=0 \mod 2$. The set of such solutions is in one-to-one correspondence with  $H^1(M_3,\Z_2)$; $s_c\in H^1(M_3,\Z_2)$ denotes the corresponding element.

\end{itemize}

Now we will focus on plumbed 3-manifolds. A plumbing graph $\Gamma$ is a graph with vertices $I\in \vert$ labelled by integers $a_I\in \Z$. By $\vert$ and $\text{Edges}$ we denote the set of all vertices and edges of $\Gamma$. We will restrict ourselves to the case when the graph is connected and there are no loops.\footnote{The generalization to graphs with loops was considered in \cite{Chun:2019mal}, where the relation between ordinary WRT invariants and the BPS $q$-series invariants $\hat Z_b (q)$ was studied. It would be interesting to extend the analysis of \cite{Chun:2019mal} to Spin-refined WRT invariants discussed here.} A plumbed 3-manifold $M_3$ corresponding to $\Gamma$ then can be constructed in the following way. Consider first a framed link $\CL(\Gamma)\subset S^3$ associated to the plumbing graph $\Gamma$, as illustrated in Figure \ref{fig:plumbing}. For each vertex $I$ one associates an unknot with framing specified by $a_I\in \Z$, its self-linking number. A presence of an edge between two vertices in $\Gamma$ indicates that the corresponding pair of unknots forms a standard Hopf link. The number of components of the link $\CL(\Gamma)$ is equal to the number of vertices in $\Gamma$, that is $L=|\vert|$. The corresponding 3-manifold $M_3$ is then obtained by a Dehn surgery on the framed link $\CL(\Gamma)$. 
The $L\times L$ linking matrix of $\CL(\Gamma)$ has the following entries:
\begin{equation}
Q_{IJ}=\left\{
\begin{array}{ll}
1,& I,J\text{ connected}, \\
a_I, & I=J, \\
0, & \text{otherwise}.
\end{array}
\right.\qquad I,J \in \vert.
\label{linking}
\end{equation}
As in the case of a general link, we will denote by $b_{\pm}$ the number of positive/negative eigenvalues of $Q$. The matrix $Q$ contains basic homotopy invariants of the 3-manifold $M_3$. In particular, the first homology group of $M_3$ is given by the cokernel of the linking matrix, understood as a linear map $Q:\Z^L\rightarrow \Z^L$:
\begin{equation}
H_1(M_3,\Z) \; \cong \; \mathrm{Coker} \, Q \; = \; \Z^L/Q\Z^L.
\label{H1-coker}
\end{equation}
Assume for simplicity that $Q$ is non-degenerate, so that $\mathrm{Coker} \, Q$ is a finite abelian group. Then $M_3$ is a rational homology sphere, {\it i.e.} $b_1(M_3)=0$. It has a natural linking pairing on the first homology group\footnote{In general, the pairing is only defined on the torsion subgroup.}
\begin{equation}
    \begin{array}{rccl}
         \lk: & H_1(M_3,\Z)\otimes H_1(M_3,\Z) & \longrightarrow & \Q/\Z, \\
         & [\gamma_1]\otimes [\gamma_2] & \longmapsto & \frac{\#(\gamma_1 \cap \beta_2)}{n}\qquad (n\gamma_2=\partial \beta_2).  
    \end{array}
    \label{linking-pairing-def}
\end{equation}
Using the isomorphism (\ref{H1-coker}), this pairing can be expressed in terms of the linking matrix:
\begin{equation}
\lk(a,b) =a^T Q^{-1} b\;\mod \Z,\qquad a,b\in  \Z^L/Q\Z^L \,.
\end{equation}
where $T$ denotes the transposition. The colored Jones polynomial of a link $\CL(\Gamma)$ colored by $n\in \Z_+^L$ reads 
\begin{multline}
J[\CL(\Gamma)]_{n}=\frac{1}{q^{1/2}-q^{-1/2}}\prod_{I\;\in\; \vert}
q^{\frac{a_I(n_I^2-1)}{4}}
\,\left(\frac{1}{q^{n_I/2}-q^{-n_I/2}}\right)^{\text{deg}(I)-1}
\times \\
\prod_{(I,J)\;\in\;\text{Edges}}
(q^{n_{I}n_{J}/2}-q^{-n_{I}n_{J}/2})
\qquad\qquad
\label{Jones-tree}
\end{multline}
where $\deg(I)=\sum_{J\neq I} Q_{IJ}$ denotes the degree of the vertex $I$. On the other hand, the BPS invariants $\hat{Z}_b (M_3,q)$ for a (weakly negative definite) plumbed $M_3$ read \cite{Gukov:2017kmk,Gukov:2019mnk}:
\begin{equation}
    \hat{Z}_b (q) \; = \; (-1)^{b_+}\,q^{\frac{3b_+ - 3b_- - \sum_I Q_{II}}{4}}\sum_{\ell\in 2Q\Z^{L}+b}F_\ell\, q^{-\frac{\ell^T Q^{-1}\ell}{4}}\qquad \in 2^{-c}q^{\Delta_b}\Z[[q]]
\label{Zhat-plumbed}
\end{equation}
where $F_\ell$ are the coefficients of the following Laurent power series
\begin{equation}
\sum_{\ell\in 2\Z^L +\delta} F_\ell\prod_{I\in \vert} x_I^{{\ell_I}/{2}}:=
\prod_{I\,\in\,\vert}\frac{1}{2}\,\left\{
 (x_I^{1/2}-x_I^{-1/2})^{2-\deg(I)} \Big|_{|x_I|<1}
+
 (x_I^{1/2}-x_I^{-1/2})^{2-\deg(I)} \Big|_{|x_I|>1}
\right\}.
\end{equation}
Note that $\hat{Z}_b = 0$ unless $b_I=\delta_I:=\deg(I)\mod 2=\sum_{J\neq I} Q_{JI} \mod 2$, so one can assume $b\in (2\Z^L+\delta)/2Q\Z^L = (2\Z^L+Q\varepsilon-\text{diag}\,(Q))/2Q\Z^L$. This subset of elements with fixed parity in $\Z^L/2Q\Z^L$ can be canonically identified with Spin$^c(M_3)$ (and, non-canonically, with $H_1(M_3,\Z)\cong \Z^L/Q\Z^L$).

To obtain a relation between $\hat{Z}_b (M_3,q)$ and $\text{WRT}(M_3,k,s)$ in the case of a (weakly negative definite) plumbed $M_3$ we will follow closely the analysis in Appendix A of \cite{Gukov:2017kmk}, where the relation to unrefined WRT invariant was obtained. We will need a slightly different version of the Gauss sum reciprocity formula than the one used in \cite{Gukov:2017kmk} (which is still a particular case of a more general formula in \cite{jeffrey1992chern,deloup2007reciprocity}):
\begin{multline}
\sum_{n\;\in\;\Z^L/k\Z^L}
\exp\left(\frac{2\pi i}{k}n^T Q n+\frac{2\pi i}{k}\ell^Tn\right)
=\\=
\frac{e^{\frac{\pi i(b_+-b_-)}{4}}\,(k/2)^{L/2}}{|\det Q|^{1/2}}
\sum_{\tilde{a}\;\in\; \Z^L/2Q\Z^L}
\exp\left(-\frac{\pi i k}{2}\left(\tilde{a}+\frac{\ell}{k}\right)^T Q^{-1} \left(\tilde{a}+\frac{\ell}{k}\right)\right)
\qquad
\label{reciprocity-alt}
\end{multline}
where $\ell \in \Z^L$ and $b_+ - b_-$ is the signature of the linking matrix $Q$. By applying (\ref{reciprocity-alt}) to (\ref{RT-sum}) with the colored Jones polynomial (\ref{Jones-tree}) we get:
\begin{multline}
    \text{WRT}(M_3,k,s_c)=
    \frac{F^{(c+\varepsilon)}[\CL(\Gamma)]}{F^\text{tot}[U_{-1}]^{b_-}F^\text{tot}[U_{+1}]^{b_+}}
    = \\ =
    \frac{1}{(\q^{1/2}-\q^{-1/2})2^{L+1}|\det Q|^{1/2}}
    \sum_{\substack{ \tilde{a}\in \Z^L/2Q\Z^L\\ \tilde{b}\in \Z^L/2Q\Z^L}} 
    e^{-\frac{\pi i k}{2}\tilde{a}^T Q^{-1} \tilde{a}}\,e^{-\pi i\tilde{a}^T Q^{-1} \tilde{b}}\,
    \hat{Z}_{\tilde{b} + Q(c+\varepsilon)} \Big|_{q\rightarrow\q}.
    \qquad    \label{spin-WRT-decomp}
\end{multline}
The terms in this decomposition are not identically zero for $\tilde{b}\in (2\Z^L+\delta-Q(c+\varepsilon))/2Q\Z^L=(2\Z^L+\text{diag}\,(Q)-Qc)/2Q\Z^L$. We are interested in the case $k=0\mod 4$. Let
$\tilde{a}={a}+QA$ where $A\in \Z_2^L$ (as an  element of a \textit{group}) and ${a}\in \Z^L/Q\Z^L$ (a priori only as an element of a \textit{set}). Then,
\begin{multline}
    \text{WRT}(M_3,k,s_c)
 = \\ =
 \frac{1}{(\q^{1/2}-\q^{-1/2})2^{L+1}|\det Q|^{1/2}} \,
    \sum_{\substack{a\in \Z^L/Q\Z^L\\A\in \Z^L/2\Z^L \\\tilde{b}\in \Z^L/2Q\Z^L}} e^{-2\pi i\frac{k}{4} \,a^TQ^{-1}a}\,e^{-\pi i\,a^TQ^{-1}\tilde{b}}\,e^{-\pi i\,A^T\tilde{b}}\,
    \hat{Z}_{\tilde{b}+Q(c+\varepsilon)} \Big|_{q\rightarrow\q}
       =  
      \\ =
      \frac{1}{(\q^{1/2}-\q^{-1/2})2|\det Q|^{1/2}}\,
    \sum_{\substack{ a\in \Z^L/Q\Z^L\\ \tilde{b}\in \Z^L/2Q\Z^L }} e^{-2\pi i \frac{k}{4} \,a^TQ^{-1}a}\,e^{-\pi i\,a^TQ^{-1}\tilde{b}}\,\delta(\tilde{b}=0\mod 2)\,
    \hat{Z}_{\tilde{b}+Q(c+\varepsilon)} \Big|_{q\rightarrow\q}
\end{multline}
where in the second line we performed explicitly the sum over $A$ to obtain a discrete delta-function condition $\tilde{b}=0\mod 2$. This condition actually does not affect the sum because, as was pointed out earlier, $\hat{Z}_{\tilde{b}+Q(c+\epsilon)} = 0$ unless $\tilde{b}+Q(c+\epsilon)=\delta\mod 2$. From the fact that $\delta_I=\sum_{I\neq J}Q_{IJ}$ and $\sum_J Q_{IJ}c_J=Q_{II}\mod 2$ it is easy to see that
\begin{equation}
    \tilde{b}+Q(c+\epsilon)=\delta\mod 2 \qquad \Longleftrightarrow \qquad \tilde{b} = 0\mod 2.
\end{equation}
Replacing $\tilde{b}=2b$, we have
$$
\text{WRT}(M_3,k,s_c)
       =   \frac{1}{(\q^{1/2}-\q^{-1/2})2|\det Q|^{1/2}} \,
    \sum_{\substack{ a\in \Z^L/Q\Z^L\\ b\in \Z^L/Q\Z^L }} e^{-2\pi i \frac{k}{4} \,a^TQ^{-1}a}\,e^{-2\pi i\,a^T Q^{-1} b}\,
\hat{Z}_{2b+Q(c+\varepsilon)} \Big|_{q\rightarrow\q} \,.
$$
The summand now is explicitly invariant under the shifts $a\rightarrow a+Q\alpha$. One can now conjecture the following formula for an arbitrary rational homology sphere $M_3$ and $k=0$ mod 4:
\begin{equation}
 \text{WRT}(M_3,k,s)       =   \frac{1}{(\q^{1/2}-\q^{-1/2})2|H_1(M_3,\Z)|^{1/2}} \,
    \sum_{a,b\in H_1(M_3,\Z)}e^{-2\pi i \frac{k}{4} \,\lk(a,a)}\,e^{-2\pi i \,\lk(a,b)}\,
    \hat{Z}_{\sigma(s,b)} \Big|_{q\rightarrow\q}
    \label{spinWRT-Zhat}
\end{equation}
where $\sigma$ is the map
\begin{equation}
    \sigma: \text{Spin}(M_3)\times H_1(M_3,\Z)\longrightarrow \text{Spin}^c(M_3)
\label{spinspincmap}
\end{equation}
producing a Spin$^c$ structure on $M_3$ from a Spin structure $s$ and $\tilde{b}\in H_1(M_3,\Z)$. It is induced by the canonical map $ B\text{Spin}\times BU(1) \rightarrow B\text{Spin}^c$  between the corresponding classifying spaces, combined with the isomorphisms $BU(1)\cong B^2\Z$, $H_1(M_3,\Z)\cong H^2(M_3,\Z)$. The map between the classifying spaces appears in the following long sequence of fibrations:
\begin{equation}
\begin{array}{ccccccl}
&\Z_2 & \longrightarrow & \text{Spin}\times U(1) & \longrightarrow  & \text{Spin}^c & \longrightarrow \\
  \longrightarrow  & B\Z_2 &\longrightarrow & B\text{Spin}\times BU(1) & \longrightarrow & B\text{Spin}^c & \longrightarrow \ldots
\end{array}
\end{equation}
When $k=4$ ({\it i.e.} $\q=i$), taking into account that  \cite{MR2457479}:
\begin{equation}
    \text{WRT}(M_3,4,s) \; = \; e^{- \frac{3\pi i\mu(M_3,s)}{8}}    
\end{equation}
the formula (\ref{spinWRT-Zhat}) gives the coefficients (\ref{cRokhlin-general}).

\begin{figure}[ht]
	\centering
	\includegraphics[width=4.0in]{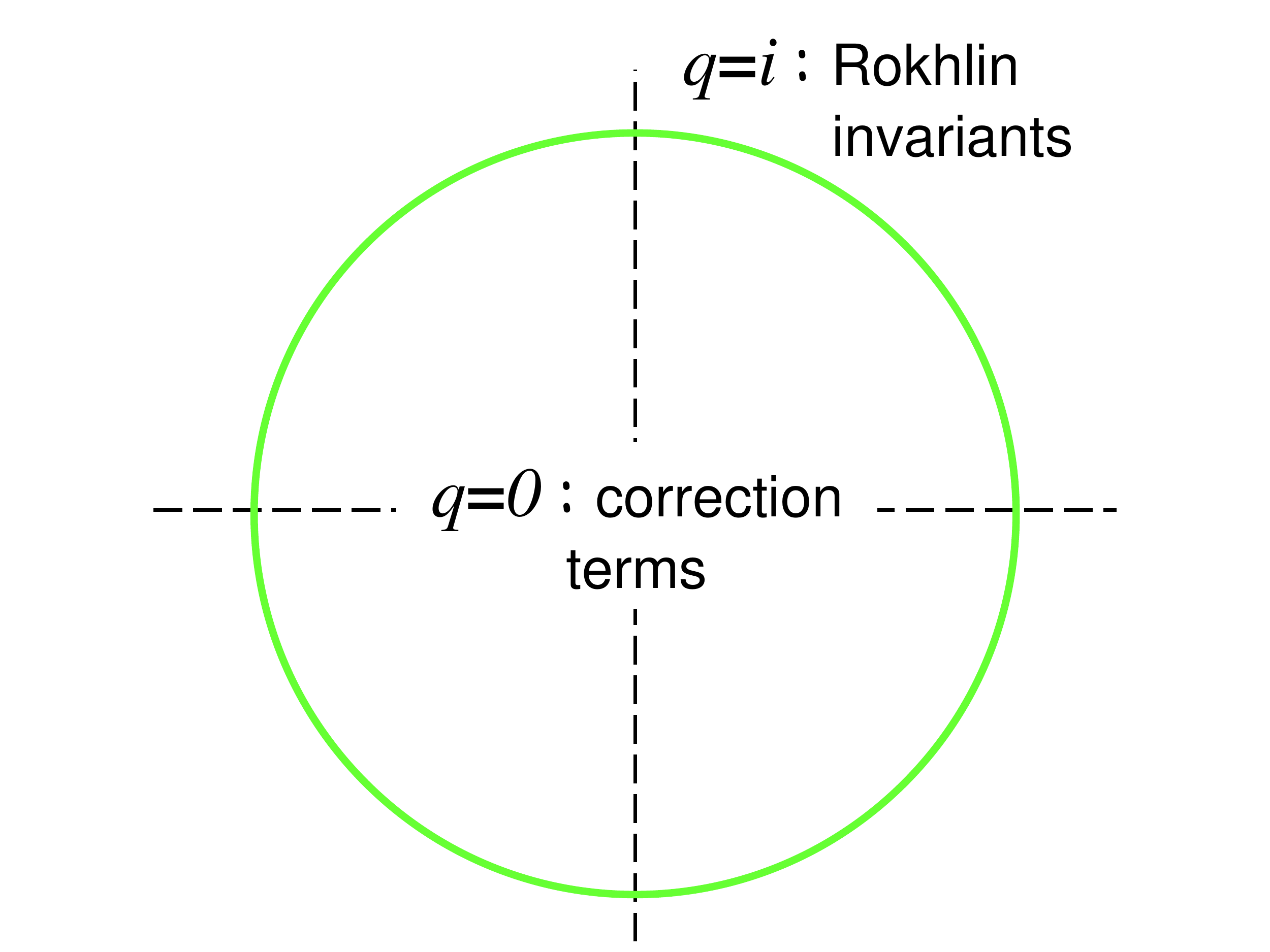}
	\caption{Various topological invariants can be found in different ``corners'' (different limits) of the BPS $q$-series.}
	\label{fig:qplane}
\end{figure}

\section{Surgeries on knots}
\label{sec:surgeries}

In this section, we explore general $\frac{p}{r}$ Dehn surgeries on knots:
\be
M_3 \; = \; S^3_{p/r} (K).
\label{generalsurgery}
\ee
In particular, we wish to study a relation between the Rokhlin invariant of $M_3$ and the behavior of the BPS $q$-series $\hat Z_b (M_3,q)$ at $q=i$, as well as a similar relation between the correction terms and the behavior of $\hat Z_b (M_3,q)$ at $q=0$, {\it cf.} Figure~\ref{fig:qplane}.

Along the way, we review various tools and techniques that can be used to compute these invariants.
For example, when $p$ is odd, {\it i.e.} \eqref{generalsurgery} admits a unique Spin structure, the Rokhlin invariant is determined by the general surgery formula\footnote{There is a typo in the last term of Theorem 2.13 in \cite{MR1941324}.} \cite{MR1941324}:
\be
\mu \left( S^3_{p/r} (K) \right)
\; = \;
\mu \left( L(p,r) \right)
+ 4r \Delta_K'' (1)
\qquad \text{mod}~16
\label{Savelievgeneral}
\ee
where $\Delta_K (x)$ is the Alexander polynomial of $K$ (not to be confused with the leading power of $q$ in \eqref{Zhatqseries}) and $\mu ( L(p,r) )$ is the Rokhlin invariants of the Lens space $L(p,r)$.
The explicit form of all these ingredients will appear further below, along with analogous surgery formulae for the correction terms and some relevant facts about the $q$-series invariants \eqref{Zhatqseries}.

\subsection{Small surgeries}

It is natural to start with the so-called small surgeries, {\it i.e.} surgeries \eqref{generalsurgery} with $|p|=1$. They all produce integral homology spheres, $H_1 (M_3; \Z) = 0$, so that $M_3$ carries a unique Spin structure and a unique Spin$^c$ structure.
In our present context, this makes such examples simpler because there is only one Rokhlin invariant $\mu (M_3)$, only one correction term $d (M_3)$,  only one $q$-series $\hat Z_0 (M_3)$ and, correspondingly, only one value of $\Delta_b (M_3)$.
In this subsection, we consider examples of such surgeries on ${\bf 3_1}$, ${\bf 4_1}$, and ${\bf 5_2}$ knots including infinite families, such as
\be
M_3 \; = \; S^3_{-1/r} ({\bf 3_1^r}).
\label{237family}
\ee

In many examples, including \eqref{237family}, the $q$-series invariants $\hat Z_b (M_3)$ are given by linear combinations of the false theta-functions
\begin{gather}
\tilde \Psi^{(a)}_p (q) \;  :=  \; \sum_{n=0}^\infty \psi^{(a)}_{2p}(n) q^{\frac{n^2}{4p}} \qquad \in q^\frac{a^2}{4p}\,\Z[[q]],
\label{falsetheta} \\
\psi^{(a)}_{2p}(n)  =  \left\{
\begin{array}{cl}
\pm 1, & n\equiv \pm a~\mod~ 2p\,, \\
0, & \text{otherwise}.
\end{array}\right. \nonumber
\end{gather}
Expressing $\hat Z_b (M_3,q)$ in terms of $\tilde \Psi^{(a)}_p (q)$ can streamline the analysis of the limiting behavior at $q=i$.
Indeed, using the well known modular properties of the false theta-functions (that can be found {\it e.g.} in \cite{MR1701924,MR2191375,Kucharski:2019fgh,Chung:2019jgw}), we have
\be
\tilde \Psi^{(a)}_p (e^{2\pi i /k})
\; = \; \frac{1}{2} \sum_{n=0}^{2pk}
\left( 1 - \frac{n}{pk} \right)
\psi_{2p}^{(a)} (n) e^{\pi i \frac{n^2}{2pk}}.
\ee
In particular, for $k=4$ this gives the desired formula
\be
\tilde \Psi^{(a)}_p (e^{2\pi i /4})
\; = \; \frac{1}{2} \sum_{n=0}^{8p}
\left( 1 - \frac{n}{4p} \right)
\psi_{2p}^{(a)} (n) e^{\pi i \frac{n^2}{8p}}
\label{falsethetalimit}
\ee
that one can now apply to any example where $\hat Z_b (M_3,q)$ can be expressed in terms of $\tilde \Psi^{(a)}_p (q)$.
This is the case for the family \eqref{237family}, which has the following $q$-series \cite{Gukov:2019mnk}:
\be
\hat Z_0 (q) \; = \; q^{\frac{1}{2} -\frac{(6r-5)^2}{24 (6r+1)}}
\left(
\tilde \Psi^{(6r-5)}_{ 6 (6r+1) } (q)
- \tilde \Psi^{(6r+7)}_{ 6 (6r+1) } (q)
- \tilde \Psi^{(30r-1)}_{ 6 (6r+1) } (q)
+ \tilde \Psi^{(30r+11)}_{ 6 (6r+1) } (q)
\right).
\ee
It is clear from \eqref{falsetheta} that in this entire family of small surgeries
\be
\Delta ( S^3_{-1/r} ({\bf 3_1^r}) ) \; = \; \frac{1}{2}.
\label{Delta237}
\ee
Moreover, \eqref{falsethetalimit} gives
\be
\frac{1}{i \sqrt{8}} \hat Z_0 (q)
\Big|_{q=i}
\; = \; (-1)^r.
\label{Z237toRokhlin}
\ee
Motivated by \eqref{muvsZ}, we wish compare these results with the values of the Rokhlin invariant and the correction terms.

For the Rokhlin invariant, we can use the general surgery formula \eqref{Savelievgeneral} that for $p=-1$ takes the form
\be
\mu ( S^3_{-1/r} (K) )
\; = \;
4r \Delta_K'' (1)
\qquad \text{mod}~16,
\label{Rokhlinsmall}
\ee
The Alexander polynomial for the trefoil knot is $\Delta_{{\bf 3_1}} (x) = x^{-1} - 1 + x$, and so $\Delta_{{\bf 3_1}}'' (1) = 2$. Therefore,
\be
\mu ( S^3_{-1/r} ({\bf 3_1^r}) )
\; = \;
\begin{cases}
8~~~(\text{mod}~16) \,, & \text{if}~ r~\text{is odd}, \\
0~~~(\text{mod}~16) \,, & \text{if}~ r~\text{is even},
\end{cases}
\label{RokhRHTsmall}
\ee
and, as expected, \eqref{Z237toRokhlin} can be written in the form \eqref{muvsZmain} with $c^{\text{Rokhlin}}$ given by (\ref{cRokhlin}a).

\begin{table}[ht]
	\begin{centering}
		\begin{tabular}{|c||c|c|c|c|}
			\hline
			$\phantom{\int^{\int^\int}} \text{Knot} \phantom{\int_{\int}}$ & ~$\sigma (K)$~ & ~$g_4 (K)$~ & ~$d (S^3_{+1} (K))$~ & ~Alternating~ \tabularnewline
			\hline
			\hline
			$\phantom{\int^{\int^\int}} {\bf 3_1^r} \phantom{\int_{\int}}$ & $-2$ & $1$ & $-2$ & \cmark 
			\tabularnewline
			\hline
			$\phantom{\int^{\int^\int}} {\bf 4_1} \phantom{\int_{\int}}$ & $0$ & $1$ & $0$ & \cmark
			\tabularnewline
			\hline
			$\phantom{\int^{\int^\int}} {\bf 5_1} \phantom{\int_{\int}}$ & $-4$ & $2$ & $-2$ & \cmark
			\tabularnewline
			\hline
			$\phantom{\int^{\int^\int}} {\bf 5_2} \phantom{\int_{\int}}$ & $-2$ & $1$ & $-2$ & \cmark
			\tabularnewline
			\hline			
		\end{tabular}
		\par\end{centering}
	\caption{\label{tab:knots} Some invariants of prime knots with up to 5 crossings.}
\end{table}

Now, let us compare the behavior of the BPS $q$-series near $q=0$, in particular the leading $q$-exponents \eqref{Delta237}, with the correction terms.
The correction terms for small surgeries can be conveniently computed using the results of \cite{MR3393360}, which say that
\be
d ( S^3_{1/r} (K) )
\; = \; - 2 V_0 (K)
\; = \; d ( S^3_{+1} (K) )
\label{dsmall}
\ee
is independent of $r$ (in the formula above, $V_0(K)$ is a certain invariant defined in terms of the knot Floer chain complex).
Moreover, for alternating knots the correction term \cite{MR1988285}:
\be
d (S^3_{+1} (K)) = 2 \, \text{min} \left( 0 , - \lceil \tfrac{- \sigma (K)}{4} \rceil \right)
\label{dalternating}
\ee
is determined by the knot signature $\sigma (K)$. For example, for the right-handed trefoil knot $\sigma ({\bf 3_1^r}) = -2$, whereas for the left-handed trefoil knot $\sigma ({\bf 3_1^{\ell}}) = 2$, {\it cf.} Table~\ref{tab:knots}. Therefore,
\begin{subequations}
\label{LRtrefcorrterms}
\be
d (S^3_{+1} ({\bf 3_1^r})) \; = \;
-2,
\ee
\be
d (S^3_{+1} ({\bf 3_1^{\ell}})) \; = \;
0.
\ee
\end{subequations}
This is consistent with the 4-ball genus bound that follows from the results of \cite{MR3393360} and \cite{MR2704683}:
\be
0 \le - \frac{d (S^3_{+1} (K))}{2} \le \Bigg\lceil \frac{g_4 (K)}{2} \Bigg\rceil.
\ee
Here, $g_4 (K)$ is the 4-ball genus of $K$ (a.k.a. the slice genus) defined as the minimal genus of the smoothly embedded surface in $B^4$ with boundary $K \subset S^3 = \partial B^4$. The values of $g_4 (K)$ for knots with small number of crossings are listed in Table~\ref{tab:knots}.

Combining \eqref{dsmall} with the behavior of the correction terms under the orientation reversal (\ref{orientation}b) we conclude
\be
d (S^3_{-1/r} ({\bf 3_1^r}))
\; = \; - d ( - S^3_{-1/r} ({\bf 3_1^r}))
\; = \; - d ( S^3_{1/r} ({\bf 3_1^{\ell}}))
\; = \; - d ( S^3_{+1} ({\bf 3_1^{\ell}}))
\; = \; 0.
\label{d237small}
\ee
This value is clearly different from \eqref{Delta237}. However, both \eqref{Delta237} and \eqref{d237small} are independent of $r$ for the entire family \eqref{237family}. We will come back to this clue shortly.

Note, for $r=1$ we can also present \eqref{237family} as a surgery on the figure-8 knot:
\be
S^3_{-1} ({\bf 3_1^r}) \; = \; S^3_{+1} ({\bf 4_1}).
\label{first41}
\ee
Therefore, it is instructive to verify \eqref{d237small} by applying \eqref{dalternating} directly to the $+1$ surgery on $K = {\bf 4_1}$.
Since $\sigma ( {\bf 4_1} ) = 0$ ({\it cf.} Table~\ref{tab:knots}), we indeed get $d (S^3_{+1} ({\bf 4_1})) = 0$.
Similarly, using the Alexander polynomial for the figure-8 knot, $\Delta_{{\bf 4_1}} (x) = - x^{-1} + 3 - x$, and the surgery formula for the Rokhlin invariant \eqref{Rokhlinsmall} we reproduce \eqref{RokhRHTsmall} in the case $r=1$, {\it i.e.} $\mu ( S^3_{+1} ({\bf 4_1}) ) = 8$ mod 16.

It is easy to generalize this analysis to small surgeries on other knots. For example, a close cousin of \eqref{237family} is
\be
M_3 \; = \; S^3_{-1/r} ({\bf 3_1^{\ell}}).
\label{235family}
\ee
The explicit form of the $q$-series $\hat Z_0 (M_3, q)$ can be found {\it e.g.} in \cite{Gukov:2019mnk}. Again, it can be written as a linear combination of the false theta-functions \eqref{falsetheta} and has
\be
\Delta ( S^3_{-1/r} ({\bf 3_1^{\ell}}) ) \; = \; - \frac{3}{2}.
\label{Delta235}
\ee
Furthermore, using \eqref{falsethetalimit} we obtain
\be
\frac{1}{i \sqrt{8}} \hat Z_0 ( S^3_{-1/r} ({\bf 3_1^{\ell}}) , q)
\Big|_{q=i }
\; = \; (-1)^r
\ee
which also confirms \eqref{muvsZmain} with (\ref{cRokhlin}a) and is the same result as we found for small surgeries on the right-handed trefoil \eqref{237family}.
The Rokhlin invariants are also the same. (This is easy to see {\it e.g.} from the surgery formula \eqref{Rokhlinsmall}.) However, the correction terms are different. Indeed, according to (\ref{LRtrefcorrterms}a), $d (S^3_{+1} ({\bf 3_1^r})) = -2$ and, therefore,
\be
d (S^3_{-1/r} ({\bf 3_1^{\ell}}))
\; = \; - d ( - S^3_{-1/r} ({\bf 3_1^{\ell}}))
\; = \; - d ( S^3_{1/r} ({\bf 3_1^r}))
\; = \; - d ( S^3_{+1} ({\bf 3_1^r}))
\; = \; 2
\label{d235small}
\ee
where the first equality is due to (\ref{orientation}b).
Again, we observe a clear parallel between \eqref{Delta235} and \eqref{d235small}; both invariants are independent of $r$ for the entire family \eqref{235family}.

Motivated by \cite{Gukov:2016gkn}, we expect a linear relation between $\Delta_b (M_3)$ and $d (M_3,b)$. Comparing \eqref{Delta237} with \eqref{d237small} and, similarly, \eqref{Delta235} with \eqref{d235small}, we see that a relation --- if it exists --- must be of the form
\be
\Delta_b (M_3) \; = \; \frac{1}{2} + \frac{d (M_3,b)}{2} n
\qquad \text{mod}~1
\label{dvsDelta}
\ee
for some $n \in \Z$. At this point, the data based on small surgeries \eqref{237family} and \eqref{235family} does not allow to determine the value of $n$ in this relation. But this will quickly change as we consider more examples; many of them will determine $n$.

Before we leave the comfort of homology spheres and small surgeries, let us consider one more example that involves a hyperbolic knot $K = {\bf 5_2}$:
\be
M_3 \; = \; S^3_{+1} ({\bf 5_2}).
\label{52plus1}
\ee
For this surgery, using the same techniques as before, we find
\be
\Delta (S^3_{+1} ({\bf 5_2})) \; = \; \frac{3}{2}
\,, \qquad
\mu (S^3_{+1} ({\bf 5_2})) \; = \; 0
\,, \qquad
d (S^3_{+1} ({\bf 5_2})) \; = \; -2
\ee
and
\be
\frac{1}{i \sqrt{8}} \hat Z_0 ( S^3_{+1} ({\bf 5_2}) , q)
\Big|_{q=i }
\; = \; 1.
\label{Zmu52plus1}
\ee
Specifically, the computation of the correction term is a direct application of \eqref{dalternating}. Similarly, the Rokhlin invariant $\mu (S^3_{+1} ({\bf 5_2})) = 0$ is determined by \eqref{Rokhlinsmall} and the Alexander polynomial $\Delta_{{\bf 5_2}} (x) = 2 x^{-1} -3 + 2x$.
The computation of \eqref{Zmu52plus1} and $\Delta (S^3_{+1} ({\bf 5_2})) = \frac{3}{2}$ is a little more interesting. We are not aware of any place in the literature where the explicit form of $\hat Z_0 ( S^3_{+1} ({\bf 5_2}) , q)$ has been worked out. However, for the manifold with opposite orientation the $q$-series has been studied in \cite{Park:2020edg} and this suffices to determine \eqref{Zmu52plus1} and $\Delta (S^3_{+1} ({\bf 5_2})) = \frac{3}{2}$ thanks to \eqref{muvsZorientation} and \eqref{Deltaorientation}, respectively. As expected, \eqref{Zmu52plus1} also supports the relation \eqref{muvsZmain} between the (exponentiated) Rokhlin invariant and the limiting value of the BPS $q$-series at $q=i$. Next, let us consider more interesting examples where dependence on Spin and Spin$^c$ structure plays an important role.

\subsection{Multiple Spin structures}

It is usually hard to think of simpler examples of knot surgeries than Lens spaces, {\it i.e.} surgeries on the unknot\footnote{It is important to pay attention to orientation conventions used in the literature. For example, since negative definite plumbings were favored in \cite{Gukov:2017kmk}, the Lens space $L(p,1)$ was naturally defined as a $-p$ surgery on the unknot. The same convention was used in \cite{MR1941324}. On the other hand, the opposite choice of orientation is used in \cite{MR1957829}, where $L(p,1)$ is a $+p$ surgery on the unknot. Here, we follow this latter choice.}
\be
M_3
\; = \; S^3_{-p} (\text{unknot})
\; = \; - L(p,1).
\label{MLens}
\ee
These simple examples, however, are quite rich as it comes to dependence on Spin and Spin$^c$ structures. In particular, as we will see shortly, a simple family \eqref{MLens} will provide infinitely many examples, each of which will uniquely determine the value of $n$ in \eqref{dvsDelta}.

For even values of $p$, \eqref{MLens} admits two Spin structures and $p$ Spin$^c$ structures, that we label by $b = 0, \ldots, p-1$. For odd values of $p$, there are $p$ Spin$^c$ structures and a unique Spin structure.
Therefore, in this class of examples, for each given $p$ we have a total of $p$ BPS invariants $\hat Z_b (q)$ and $p$ correction terms. Among $\{ \hat Z_b (q) \}_{b = 0, \ldots, p-1}$ only two are non-zero and the expressions that are sometimes quoted, $\hat Z_0 (q) = -2$ and $\hat Z_1 (q) = 2q^{1/p}$, for convenience omit\footnote{In order to fully account for BPS states in 3d $\CN=2$ theory with 2d $(0,2)$ boundary condition or in relation to vertex algebras \cite{Cheng:2018vpl} one also needs to restore factors of $(q;q)_{\infty}$ that correspond to the ``center-of-mass'' chiral multiplet in 3d $\CN=2$ theory and are also usually omitted. The normalization of $\hat Z_b (M_3,q)$ that relates to topological invariants of $M_3$, such as WRT invariants, never includes such factors, though, and so they will not play any role here.} the overall factor $q^{\frac{p-3}{4}}$. For our applications here, it is important to not separate any such factors and, therefore,
\be
\hat Z_0 (q) \; = \; - 2q^{\frac{p-3}{4}}
\,, \qquad
\hat Z_1 (q) \; = \; 2q^{\frac{p^2-3p+4}{4p}}.
\label{ZLens}
\ee
In particular, plugging these expressions into \eqref{WRTfromZhat}, it is easy to check that it gives the correct values of WRT invariants for \eqref{MLens} for all values of $p$ and $k$. Moreover, from \eqref{ZLens} it is clear that
\be
\Delta_0 \; = \; \frac{p-3}{4}
\,, \qquad
\Delta_1 \; = \; \frac{p^2-3p+4}{4p},
\label{DeltaLens}
\ee
which we compare with the correction terms of Lens spaces shortly.

In order to compare \eqref{ZLens} at $q=i$ with the Rokhlin invariants, we substitute these $q$-series invariants --- which, in this class of examples, are simply monomials --- into the proposed relation \eqref{muvsZmain} and find
\be
\sum_{b=0}^{p-1}
c^{\text{Rokhlin}}_{s,b} \; \hat Z_b (q) \Big|_{q=i} \; = \; 
\begin{cases}
e^{- \frac{3 \pi i}{8} (p-1)} \,,  & \text{if}~ p~\text{is odd}, \\
e^{- \frac{3 \pi i}{8} (p-1)}~\text{and}~e^{\frac{3 \pi i}{8}} \,, & \text{if}~ p~\text{is even},
\end{cases}
\ee
where the explicit form of $c^{\text{Rokhlin}}_{s,b}$ is given in (\ref{cRokhlin}b) and (\ref{cRokhlin}c), respectively.
In other words, we find
\be
\mu (M_3,s) \; = \;
\begin{cases}
p-1 \,,  & \text{if}~ p~\text{is odd}, \\
p-1~\text{and}~-1 \,, & \text{if}~ p~\text{is even},
\end{cases}
\label{muLens}
\ee
which are indeed the correct values of the Rokhlin invariants for Lens spaces \eqref{MLens}.
One way to see this is to compute the Neumann-Siebenmann invariant $\bar \mu (M_3,s) \in \Z$ that provides an integral lift of the Rokhlin invariant \cite{MR585657}:
\be
\mu (M_3,s) \; = \; \bar \mu (M_3,s)
\qquad \text{mod}~16
\label{NSvsRokhlin}
\ee
and is defined for any plumbed 3-manifold $M_3 = M_3 (\Gamma)$, {\it cf.} \eqref{muviacharsurf}:
\be
\bar \mu (M_3 (\Gamma),s) \; = \;
\sigma (M_4 (\Gamma)) - w \cdot w \; \in \; \Z.
\label{mubardef}
\ee
Here, $M_4 (\Gamma)$ is a simply-connected 4-manifold defined by the same plumbing tree $\Gamma$ as $M_3 (\Gamma)$, and $w = w (\Gamma, s) \in H_2 (M_4 (\Gamma),\Z)$ is the spherical Wu class associated with Spin structure $s$. In particular, \eqref{charcondition} holds with $w = [\Sigma]$. Note, although a given $M_3$ can be realized by many different plumbing graphs and the choice of $w (\Gamma,s)$ is not unique, the invariant \eqref{mubardef} is well defined and, moreover, gives a Spin $\Z$-homology cobordism invariant.
Since our examples \eqref{MLens} can be realized by a graph $\Gamma$ with a single vertex labeled by $-p$, the signature of $M_4 (\Gamma)$ is equal to $-1$ and we quickly obtain $\bar \mu = p-1$ for $p$ odd, and $\bar \mu = \{ p-1 , -1 \}$ for $p$ even, in complete agreement with its mod 16 reduction quoted in \eqref{muLens}.

Now let us compare \eqref{DeltaLens} with the correction terms for the same manifolds \eqref{MLens}. For $L(p,1)$ defined as $+p$ surgery on the unknot, the correction terms $d (L(p,1),b)$ are given by \cite{MR1957829}:
\be
d (L(p,1),b) \; = \; \frac{(p-2b)^2 - p}{4p}
\,, \qquad b = 0, 1, \ldots, p-1.
\label{Lenscorrterms}
\ee
For convenience, here we list the first few values that will be also helpful to us in other examples:
\be
\begin{array}{l@{\;}|@{\;}ccccccc}
L(p,1) & \multicolumn{7}{c}{~~~~~\text{correction terms}~~~~~} \\ [.1cm]
\hline
L(2,1) \phantom{\oint^{\oint}} & ~ & \frac{1}{4} & -\frac{1}{4} & ~ & ~ & ~ & ~ \\ [.2cm]
L(3,1) & ~ & \frac{1}{2} & -\frac{1}{6} & -\frac{1}{6} & ~ & ~ & ~ \\ [.2cm]
L(4,1) & ~ & \frac{3}{4} & 0 & -\frac{1}{4} & 0 & ~ & ~ \\ [.2cm]
L(5,1) & ~ & 1 & \frac{1}{5} & -\frac{1}{5} & -\frac{1}{5} & \frac{1}{5} & ~ \\ [.2cm]
L(6,1) & ~ & \frac{5}{4} & \frac{5}{12} & -\frac{1}{12} & -\frac{1}{4} & -\frac{1}{12} & \frac{5}{12} \\ [.2cm]
~~\vdots & \multicolumn{7}{c}{~~~~~\cdots~~~~~}
\end{array}
\label{Lcorrtermssmallp}
\ee
Since under the orientation reversal on a 3-manifold the correction terms change sign, for our examples \eqref{MLens} we obtain
\be
d (-L(p,1),0) \; = \; - \frac{p-1}{4}
\,, \qquad
d (-L(p,1),1) \; = \; - \frac{p^2 - 5p + 4}{4p}.
\ee
Comparing these values with \eqref{DeltaLens}, we see that, for every $p = 2, 3, \ldots $, the following relations hold:
\begin{subequations}
\label{dvsDeltaLens}
\be
\Delta_0 (M_3) \; = \; - \frac{1}{2} - d (M_3,0),
\ee
\be
\Delta_1 (M_3) \; = \; + \frac{1}{2} - d (M_3,1).
\ee
\end{subequations}
In particular, each value of $p$ can be used to determine $n=-2$ in \eqref{dvsDelta}.

It is interesting to generalize this analysis to other Lens spaces,
\be
M_3
\; = \; S^3_{-p/r} (\text{unknot})
\; = \; - L(p,r)
\ee
with $r>1$. They also can be represented by plumbing graphs with vertices labeled by coefficients in the continued fraction expansion of $- \frac{p}{r}$ and, therefore, \eqref{NSvsRokhlin}--\eqref{mubardef} can be used to compute the Rokhlin invariants $\mu (M_3,s)$.
The correction terms of $L(p,r)$ can be expressed in terms of the Dedekind sums
\bea
s (h,k) & = & \sum_{r=1}^{k-1} \frac{r}{k} \left(
\frac{hr}{k} - \Big\lfloor \frac{hr}{k} \Big\rfloor - \frac{1}{2} \right) \notag \\
& = & \sum_{r=1}^{k-1}
\Big( \Big( \frac{r}{k} \Big) \Big)
\Big( \Big( \frac{hr}{k} \Big) \Big) \label{Dedekindsum} \\
& = & \frac{1}{4k} \sum_{r=1}^{k-1}
\cot \left( \frac{\pi h r}{k} \right)
\cot \left( \frac{\pi r}{k} \right)
\notag
\eea
where
\be
((x)) \; = \;
\begin{cases}
	x - \lfloor x \rfloor - \frac{1}{2} \,, & \text{if}~ x\not\in \Z, \\
	0 \,, & \text{if}~ x \in \Z.
\end{cases}
\ee
Specifically \cite{MR2461872}:
\be
d (L(p,r),b) \; = \;
3 s (r,p) + \frac{p-1}{2p} + 2 \sum_{k=1}^b \left( \Big( \frac{2r' k - 1}{2p} \Big) \right)
\label{dprbLens}
\ee
where $r' \in \{ 1, \ldots, p-1 \}$ is uniquely determined by $rr' \equiv 1$ (mod p).
Note, averaging the correction terms over all Spin$^c$ structures gives \cite{MR2087076} the Casson-Walker invariant of $L(p,r)$:
\be
\lambda (L(p,r)) \; = \; \frac{1}{p} \sum_{b \in \text{Spin}^c (L(p,r))} d (L(p,r),b).
\ee
It would be interesting to interpret this sum over Spin$^c$ structures in terms of BPS $q$-series invariants, as a relation analogous to \eqref{inverseTuraev}--\eqref{muvsZ}.

\subsection{Integral surgeries}

The next natural step is to consider integral surgeries on more general knots. For instance,
\be
M_3 \; = \; S^3_{-3} ({\bf 3_1^r})
\label{int31surgery}
\ee
combines several aspects of the previous examples \eqref{237family} and \eqref{MLens}. Namely, it has $H_1 (M_3;\Z) = \Z_3$ and, much like \eqref{MLens}, has collections of invariants $\{ \hat Z_b (q) \}$, $\{ \Delta_b \}$ and $\{ d(b) \}$, labeled by the Spin$^c$ structure~$b$.
On the other hand, each $\hat Z_b (q)$ is a non-trivial $q$-series, so that computing its limiting values and comparing them with Rokhlin invariants requires some work, as in \eqref{237family}.

For the 3-manifold \eqref{int31surgery}, the BPS $q$-series are given by \cite{Cheng:2018vpl,Gukov:2019mnk}:
\begin{subequations}
\label{Zhatint31}
\be
\hat Z_0 = q^{\frac{71}{72}}
\left( \tilde \Psi_{18}^{(1)} + \tilde \Psi_{18}^{(17)}  \right)
= q + q^5 - q^6 - q^{18} + q^{20} + \ldots,
\ee
\be
\hat Z_1 = - q^{\frac{71}{72}}
\left( \tilde \Psi_{18}^{(5)} + \tilde \Psi_{18}^{(13)}  \right)
= - q^{4/3} \left( 1 + q^2 - q^7 - q^{13} + q^{23} + \ldots \right),
\ee
\end{subequations}
where $\tilde \Psi_{p}^{(a)}$ are the familiar false theta-functions \eqref{falsetheta}.
In particular, from the leading $q$-powers we infer
\be
\Delta_0 = 1
\,, \qquad
\Delta_1 = \frac{4}{3}.
\label{Delta31int}
\ee
And, using \eqref{falsethetalimit}, we find the limiting values at $q=i$, which then can be plugged into \eqref{muvsZmain} to give
\be
\sum_{b}
c^{\text{Rokhlin}}_{b} \; \hat Z_b (q) \Big|_{q=i}
\; = \;
e^{ - 2 \pi i \frac{7}{8}}
\ee
with the coefficients $c^{\text{Rokhlin}}_{b}$ written in (\ref{cRokhlin}b).
This result is in perfect agreement with the value of the Rokhlin invariant
\be
\mu (S^3_{-3} ({\bf 3_1^r})) \; = \; 10
\qquad \text{mod}~16
\label{Rokhlin31int}
\ee
that can be computed by several techniques mentioned earlier. First, we can use the general surgery formula \eqref{Savelievgeneral}, which in this case gives
\be
\mu ( S^3_{-3} ({\bf 3_1^r}) )
\; = \;
\mu \left( - L(3,1) \right)
+ 4 \Delta_{{\bf 3_1^r}}'' (1) \; = \; 10
\qquad \text{mod}~16.
\ee
The first term here comes from \eqref{muLens} and the second term already appeared in \eqref{Rokhlinsmall}.
We can also compute \eqref{Rokhlin31int} by representing \eqref{int31surgery} as a plumbed manifold, {\it e.g.} with the following plumbing graph \cite{Cheng:2018vpl}:
\begin{equation}
\begin{array}{ccc}
& \overset{\displaystyle{-3}}{\bullet} & \\
& \vline & \\
\overset{\displaystyle{-2}}{\bullet}
\frac{\phantom{xxx}}{\phantom{xxx}}
& \underset{\displaystyle{-1}}{\bullet} &
\frac{\phantom{xxx}}{\phantom{xxx}}
\overset{\displaystyle{-9}}{\bullet}
\end{array}
\label{plumbing31int}
\end{equation}
It has a negative definite adjacency matrix $Q$ of size $4 \times 4$ and the Wu set $S$ that includes three outer vertices of the plumbing graph \eqref{plumbing31int} but not the central vertex.\footnote{Recall, $S$ consists of all vertices $I \in \text{Vert} (\Gamma)$ that satisfy
\be
\sum_{I \in S} Q_{IJ} \; \equiv \; Q_{JJ}
\qquad \text{mod}~2
\ee
for any $J \in \text{Vert} (\Gamma)$. This is nothing but the familiar condition \eqref{charcondition} expressed in terms of the adjacency matrix of the plumbing graph. The corresponding Kirby diagram is obtained by replacing every vertex of the graph $\Gamma$ by a copy of the unknot.} Therefore, its Neumann-Siebenmann invariant is
\bea
\bar \mu ( M_3 ) & = & \text{sign} \, Q - \sum_{I \in S} Q_{II} \\
& = & -4 - (-2-3-9) \; = \; 10
\eea
and its mod 16 reduction gives the Rokhlin invariant \eqref{Rokhlin31int}.

Now let us compare \eqref{Delta31int} with the values of the correction terms. Since $S^3_{-3} ({\bf 3_1^r}) = - S^3_{+3} ({\bf 3_1^{\ell}})$, according to (\ref{orientation}b) we have
\be
d (S^3_{-3} ({\bf 3_1^r}) , b)
\; = \; - d (S^3_{+3} ({\bf 3_1^{\ell}}) , b).
\ee
On the other hand, in our previous discussion we already had to compute $d (S^3_{+1} ({\bf 3_1^{\ell}})) = 0$, {\it cf.} (\ref{LRtrefcorrterms}b).
Then, from the results of \cite{MR3393360} it follows that the correction terms of $S^3_{+3} ({\bf 3_1^{\ell}})$ are equal to those of $L(3,1)$. In the special case of Spin$^c$ structure $b = 0$, this can be also seen from the general integral surgery formula on a knot $K$:
\be
d (S^3_{p} (K) , 0) \; = \; d (L(p,1)) + d (S^3_{+1} (K)).
\ee
The correction terms for Lens spaces already appeared in \eqref{Lenscorrterms}; in particular, for $L(3,1)$ the explicit values were listed in \eqref{Lcorrtermssmallp}.
Therefore, in our example \eqref{int31surgery} we obtain
\be
d ( S^3_{-3} ({\bf 3_1^r}) , b )
\; = \; - d (L(3,1))
\; = \; \Big\{ - \frac{1}{2} , \; \frac{1}{6} , \; \frac{1}{6} \Big\}.
\ee
Comparing these with \eqref{Delta31int}, we conclude, {\it cf.} \eqref{dvsDeltaLens}:
\begin{subequations}
\label{dvsDelta31int}
\be
\Delta_0 (M_3) \; = \; \frac{1}{2} - d (M_3,0),
\ee
\be
\Delta_1 (M_3) \; = \; \frac{3}{2} - d (M_3,1),
\ee
\end{subequations}
which again uniquely determines $n=-2$ in the provisional relation \eqref{dvsDelta} and, hence, supports \eqref{dvsDeltamain}.
We also studied several other examples of integral surgeries, similar to \eqref{int31surgery}, and in all cases found that the relations \eqref{dvsDeltamain} and \eqref{muvsZmain} hold.

Following the discussion in section \ref{sec:spin-WRT-plumbed}, we can generalize the simple example \eqref{plumbing31int} to arbitrary negative definite plumbings and show that the relation \eqref{dvsDeltamain} between $\hat Z$-invariants and the correction terms holds in this entire class.
Indeed, according to \cite{MR1988284,MR2140997}, for plumbings with $b_+ = 0$,
\be
d(M_3,b) \; = \; \frac{L + w^T Q^{-1} w}{4} \mod 1
\label{d-plumbed}
\ee
where $w \in \mathrm{Char} \, (Q) := \{w \in \Z^L \; | \; w^T n = n^T Q n \mod 2, \; \forall n \in \Z^L \}$ is a characteristic vector associated to a Spin$^c$ structure $b$ via
\begin{multline}
\text{Spin}^c (M_3) \cong \{b \in \Z^L /2Q\Z^L \; | \; b_I = \deg(I) \mod 2\} \cong
\\ \cong  \mathrm{Char} \, (Q) / 2Q \Z^L =
    \{
    w \in \Z^L / 2Q \Z^L \; | \; w_I=Q_{II}\mod 2
    \}.
\qquad
\label{plumbed-spinc}
\end{multline}
On the other hand, from \eqref{Zhat-plumbed} we find
\be
\Delta_b (M_3) \; = \; \frac{3b_+ - 3b_- -\Tr Q}{4} - \frac{\ell^T Q^{-1} \ell}{4} \mod 1
\label{Delta-plumbed}
\ee
for any $\ell = 2 \tilde b + Q(s-\varepsilon) \mod 2 Q \Z^L$ and $b=\sigma(s,\tilde b)$. Under the identification (\ref{plumbed-spinc}) we have $\ell = w - Q \epsilon \mod 2Q \Z^L$. Plugging this into \eqref{Delta-plumbed} and using $w^T \varepsilon = \Tr Q \mod 2$ along with $\sum_{I \neq J} Q_{IJ} = 2L - 2$ we obtain
\be
\Delta_b (M_3) \; = \; \frac{1}{2} - \frac{L}{4} - \frac{w^T Q^{-1} w}{4} - \frac{b_+}{2} \mod 1.
\ee
Comparing this to \eqref{d-plumbed}, we see that \eqref{dvsDeltamain} indeed holds for negative definite plumbings.

\subsection{Hyperbolic surgeries}
\label{sec:hyperbolic}

There are way more hyperbolic knots than non-hyperbolic knots and, correspondingly, there are many more hyperbolic 3-manifolds than non-hyperbolic 3-manifolds for which $q$-series $\hat Z_b (M_3,q)$ are available.
These are obtained by applying the surgery formula \cite{Gukov:2019mnk} to recent results on knot complements \cite{Park:2020edg} that provide the general tools and also perform explicit computations for knots with 7, 8, 9, and even 10 crossings.

While producing the explicit form of the $q$-expansion is no longer a challenge for many hyperbolic 3-manifolds, the task of computing the limiting values at $q=i$ is highly non-trivial and we are not aware of any general methods that can aid such computations. It would certainly help to understand modular properties of $\hat Z_b (M_3,q)$ along the lines of \cite{Cheng:2018vpl,MR4015713,Bringmann:2018ddv,MR4055703}. Until such general tools are developed, we resort to numerical methods.

We consider one of the simplest hyperbolic surgeries:
\be
M_3 \; = \; S^3_{-1/2} ({\bf 4_1})
\label{41half}
\ee
and perform the simplest form of the analysis which requires neither advanced numerical methods nor computation times that on a modest laptop take longer than a typical coffee break.
The coefficients of the BPS $q$-series can be recursively generated from the $q$-difference equation obtained by quantizing the $A$-polynomial curve of the figure-8 knot \cite{Gukov:2019mnk}:
\begin{multline}
\hat Z (S^3_{-1/2} ({\bf 4_1}))
\; = \; - q^{-\frac{1}{2}}
\big( 1-q+2 q^3-2 q^6+q^9+3 q^{10}+q^{11}-q^{14}-3 q^{15}-q^{16}+2 q^{19}+2 q^{20}
+ \ldots \\
\ldots -15040 q^{500} + \ldots \big).
\label{Z41half}
\end{multline}
And, using the techniques described above, we find the other relevant invariants:
\be
\Delta (S^3_{-1/2} ({\bf 4_1})) \; = \; - \frac{1}{2}
\,, \qquad
\mu (S^3_{-1/2} ({\bf 4_1})) \; = \; 0
\,, \qquad
d (S^3_{-1/2} ({\bf 4_1})) \; = \; 0.
\ee
Indeed, the value of $\Delta (S^3_{-1/2} ({\bf 4_1}))$ follows directly from \eqref{Z41half}. The Rokhlin invariant $\mu (S^3_{-1/2} ({\bf 4_1}))$ was basically computed below eq. \eqref{first41}, when combined with \eqref{Rokhlinsmall} for $r=2$. Similarly, the value of the correction term $d (S^3_{-1/2} ({\bf 4_1}))$ can be easily obtained from \eqref{dsmall}--\eqref{dalternating}, as in many previous examples.
Using these values, we see that the relation \eqref{dvsDeltamain} indeed holds.

\begin{figure}[h]
	\centering
	\begin{minipage}{0.45\textwidth}
		\centering
		\includegraphics[width=1.0\textwidth]{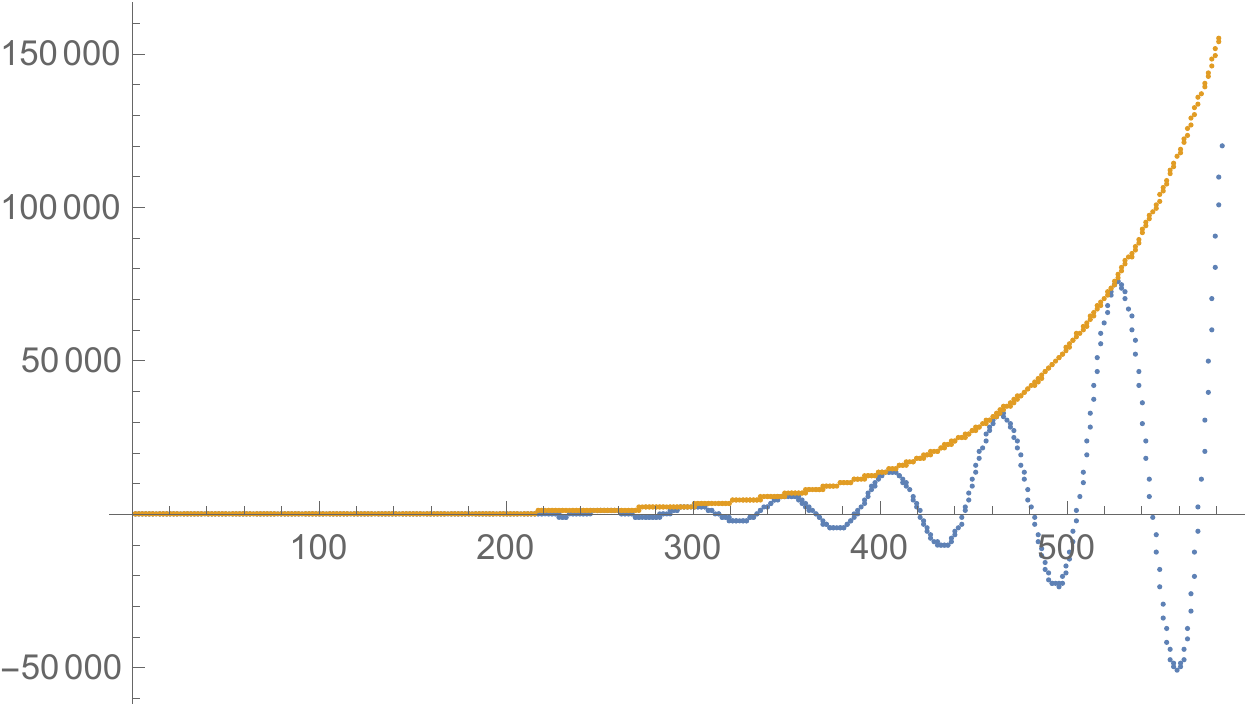}
		\notag
		\label{fig:left}
	\end{minipage}
	\qquad
	\begin{minipage}{0.45\textwidth}
		\centering
		\includegraphics[width=1.0\textwidth]{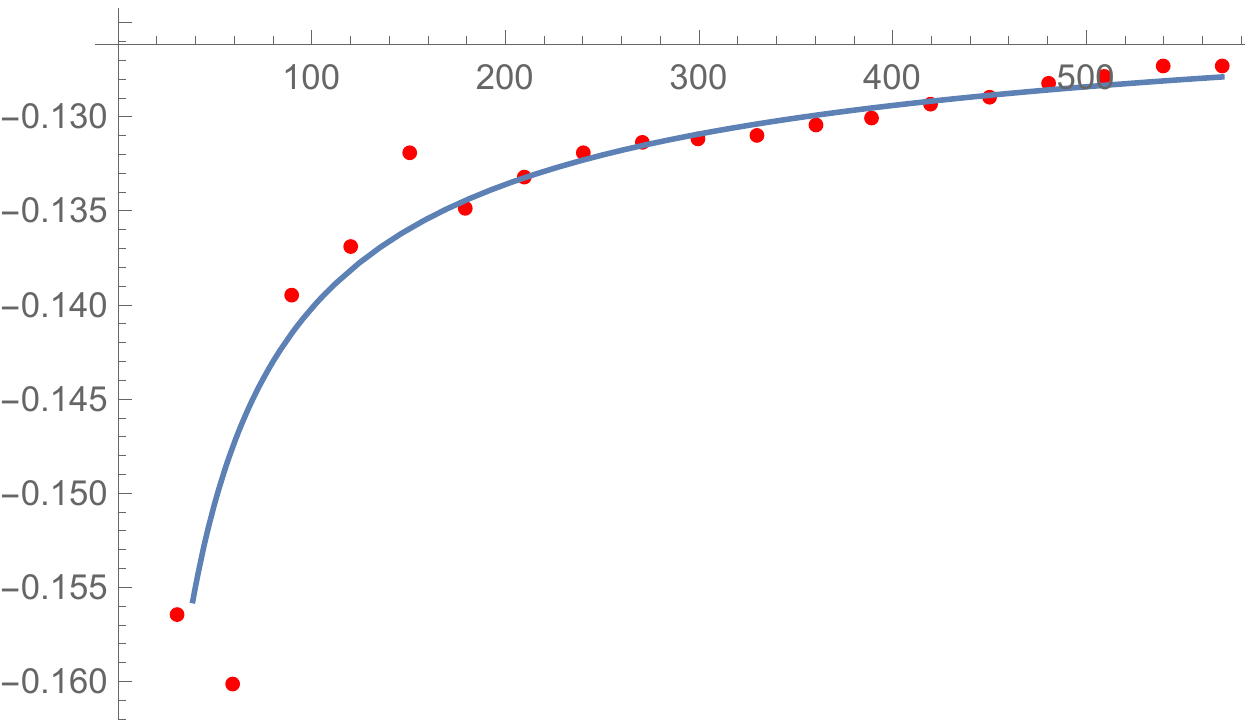}
		\notag
		\label{fig:right}
	\end{minipage}
	\caption{(Left) The coefficients of the $q$-series $-q^{\frac{1}{2}} \, \hat Z (S^3_{-1/2} ({\bf 4_1}))$ oscillate with a growing amplitude. (Right) The limiting value of $\frac{1}{2\pi} \arg (-q^{\frac{1}{2}} \, \hat Z (S^3_{-1/2} ({\bf 4_1})))$ at $q=i$ computed by approximating the $q$-series with a finite number of terms, using the prescription described in the main text. Including larger number of terms improves the accuracy, so that the sequence of approximations is approaching the expected value $- 0.125$.}
	\label{fig:half41}
\end{figure}

In order to verify \eqref{muvsZmain}, we need to compute the limiting value of \eqref{Z41half} as $q \to i$ radially, inside the unit disk $|q| < 1$. In other words, with $q = e^{2\pi i \left( \frac{1}{4} + iy \right)}$ and $y \in \R_+$, we are interested in the $y \to 0$ limit of the following quantity
\be
\frac{1}{2\pi} \arg \hat Z (S^3_{-1/2} ({\bf 4_1}))
\; = \; \frac{3}{8} +
\frac{1}{2\pi} \arg \, (-q^{\frac{1}{2}} \, \hat Z (S^3_{-1/2} ({\bf 4_1})))
\label{phase41half}
\ee
where $\frac{3}{8}$ on the right-hand side is the contribution to the phase from the overall factor $- q^{- \frac{1}{2}}$ in \eqref{Z41half}, whereas the second term is the contribution of the $q$-series $-q^{\frac{1}{2}} \, \hat Z (S^3_{-1/2} ({\bf 4_1})) = 1 - q + 2 q^3 + \ldots$. It is this latter contribution that we need to estimate numerically. First, we note that the coefficients of this $q$-series exhibit oscillatory behavior with a slowly decreasing frequency and a fast growing amplitude. This can be seen already from the first few terms in \eqref{Z41half} and is also illustrated in the left panel of Figure~\ref{fig:half41}, where the coefficients of $q^{\frac{1}{2}} \, \hat Z (S^3_{-1/2} ({\bf 4_1}))$ are shown in blue. The yellow curve, $\exp \left( 0.25 \, n^{0.61} \right)$, shows the approximate behavior of the amplitude of these oscillations. This determines the optimal truncation of the series \eqref{Z41half} to be roughly at $q^n$ with $n \approx 1020 - 128600 y$, at least in the range $0.003 \leq y \leq 0.007$ relevant to the first few hundred of terms. The corresponding data points, shown in intervals of 30 on the right panel of Figure~\ref{fig:half41}, approach the expected value $-\frac{1}{8}$ roughly as $-0.122 - 0.362 n^{-0.65}$. Indeed, with the second term contributing $- \frac{1}{8}$ to the right-hand side of \eqref{phase41half} we have
\be
\frac{1}{2\pi} \arg \hat Z (S^3_{-1/2} ({\bf 4_1}))
\Big|_{q=i}
\; = \; \frac{1}{4}
\ee
and
\be
\frac{1}{i \sqrt{8}} \hat Z ( S^3_{-1/2} ({\bf 4_1}))
\Big|_{q=i}
\; = \; 1
\ee
in agreement with \eqref{muvsZmain} and $\mu (S^3_{-1/2} ({\bf 4_1})) = 0$ obtained earlier. (Although we mainly focused on the phase of $\hat Z (S^3_{-1/2} ({\bf 4_1}))$, which gives the exponentiated Rokhlin invariant, it is easy to check that the absolute value also has the expected behavior, consistent with \eqref{muvsZmain} and (\ref{cRokhlin}a)).

\subsection{Implications for $F_K (x,q)$}

Now we discuss certain general aspects of the surgery formula and its implications for $\hat Z$-invariants of knot complements.

In general, for a knot complement $M_3 = S^3 \setminus K$, the invariant $\hat Z (S^3 \setminus K) = \{ f_{m,n,b} (q) \}$ is a collection of $q$-series invariants labeled by a pair\footnote{The invariants $\hat Z_b (M_3,q)$ provide a non-perturbative definition of $SL(2,\C)$ Chern-Simons theory (that behaves well under cutting and gluing) since the perturbative expansion of the latter is reproduced in the limit $\hbar \to 0$ (with $q = e^{\hbar}$). And, $\CH (T^2)$ can be thought of as the Hilbert space in this theory \cite{Gukov:2019mnk}, produced by quantizing the classical phase space $\CM_{\text{flat}} (T^2, SL(2,\C)) \cong \frac{\C^* \times \C^*}{\Z_2}$.} of integers $(m,n) \in \frac{\Z \times \Z}{\Z_2} = \CH (T^2)$ and a relative Spin$^c$ structure $b \in \Z \cong \text{Spin} (M_3, \partial M_3)$. When $K$ is a knot with zero framing, the action of symmetries~\cite{Gukov:2019mnk} --- which include the mapping class group of $T^2 = \partial M_3$ --- allow to set $n=0$ and $b=0$, so that the invariant of a knot complement can be written in a more compact form as
\be
F_K (x,q) \; := \; \hat Z \left( S^3 \setminus K \right)
\; = \; \sum_{m=1}^{\infty} (x^{\frac{m}{2}} - x^{- \frac{m}{2}}) \, f_m (q).
\label{FKcomplement}
\ee
Provided a collection of $q$-series invariants, conveniently packaged into $F_K (x,q)$, is known for a knot $K$ one can obtain BPS $q$-series invariants for surgries on $K$ by applying the surgery formula to $F_K (x,q)$:
\begin{equation}
\label{eq:GMsurgery}
\hat{Z}_b \left( S^3_{p/r}(K) \right) \; = \; \epsilon q^{\Upsilon} \mathcal{L}^{(b)}_{p/r} \left[ (x^{\frac{1}{2r}}-x^{-\frac{1}{2r}})F_K(x,q) \right]
\end{equation}
whenever the right-hand side makes sense (see \cite{Gukov:2019mnk} for details).
Here, $\epsilon\in \{\pm 1\}$ and $\Upsilon \in \mathbb{Q}$ a priori can depend on the knot $K$, on the surgery coefficient $p/r$, as well as on the Spin$^c$ structure $b$.
Yet, our first claim is that the exponent of the overall $q$-factor in \eqref{eq:GMsurgery} is given by the following universal formula
\begin{equation}
\label{eq:qdterm}
\Upsilon \; = \; \frac{3 (1-4s(p,r))}{4} \,\text{sign}(p)
- \frac{p}{4r} \,,
\end{equation}
where $s(p,r)$ is the Dedekind sum \eqref{Dedekindsum}.
In particular, $\Upsilon$ is independent of the knot $K$.

\begin{figure}[ht]
	\centering
	\includegraphics[scale=0.8]{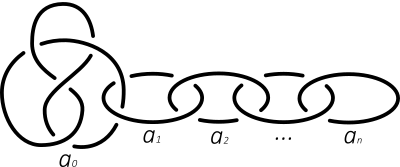}
	\caption{A rational surgery on a knot $K$ can be equivalently represented by an integral surgery on a link $L$ that consists of $K$ linked to a chain of unknots.}
	\label{fig:contfract}
\end{figure}

In order to show \eqref{eq:qdterm} it is convenient to use an equivalent\footnote{This equivalence can be shown by 3d Kirby moves.} presentation of $S^3_{p/r} (K)$ as an {\it integral} surgery on a link $L$ that consists of the original knot $K$ and a chain of unknots, shown in Figure~\ref{fig:contfract}.
The integer coefficients $a_I$ are related to $\frac{p}{r}$ via the continued fraction expansion:
\be
\frac{p}{r} \; = \; a_0 - \frac{1}{a_1 - \frac{1}{a_2 - \frac{1}{\cdots - \frac{1}{a_n}}}}.
\ee
Now, we can apply the integral surgery formula, which has a better understood overall $q$-power ({\it cf.} \cite{Gukov:2017kmk,Gukov:2019mnk} and section \ref{sec:spin-WRT-plumbed} above). Namely, for a link $L$ shown in Figure~\ref{fig:contfract} with linking matrix $Q$, it has the form
\be
\hat{Z}_b(S^3(L)) \; = \; q^{\frac{3\sigma(Q) - \Tr Q}{4}}\mathcal{L}^{(b)}_{Q} \left[
F_L(x_0, x_1, \ldots, x_n, q)
\prod_{I=0}^{n}(x_I^{\frac{1}{2}}-x_I^{-\frac{1}{2}}) 
\right]
\ee
up to a sign. Therefore, from $S^3_{p/r}(K) = S^3(L)$ it follows
\be
\hat{Z}_b(S^3_{p/r}(K)) \; = \; q^{\frac{3\sigma(Q) - \Tr Q}{4}}\mathcal{L}^{(b)}_{Q}
\left[ (x_n^{\frac{1}{2}} - x_n^{-\frac{1}{2}}) F_K(x_0,q) \right].
\ee
Now, by carefully eliminating the variable $x_n$ we obtain the rational surgery formula \eqref{eq:GMsurgery}. Note, since the explicit form of $F_K (x,q)$ did not play a role in this argument, it follows that the overall factor $q^{\Upsilon}$ in \eqref{eq:GMsurgery} is universal and does not depend on the knot $K$. Hence, \eqref{eq:qdterm} can be easily determined by taking $K$ to be the unknot.

Returning to the main theme of this paper, it is natural to ask about implications of \eqref{eq:GMsurgery}--\eqref{eq:qdterm} for the invariants $\Delta_b$ and their relation to the correction terms \eqref{dvsDeltamain}. According to Proposition 1.6 in \cite{MR3393360}, the correction terms $d (S^3_{p/r} (K),b)$ mod 2 are actually independent of $K$. Therefore, by taking $K$ to be the unknot, and using \eqref{dprbLens} we see that $d (S^3_{p/r} (K),b)$ mod 1 also admits an expression in terms of Dedekind sums, similar to \eqref{eq:qdterm}. In fact, by comparing these two expressions we observe that the proposed relation \eqref{dvsDeltamain} would hold in this class of examples if $F_K (x,q)$ (or, equivalently, all of $f_m (q)$) had only integer powers of $q$.
Therefore, a mutual consistency of these relations suggests that, unlike $\hat Z$-invariants for closed 3-manifolds, the corresponding invariants for knot complements \eqref{FKcomplement} should not have overall factors $q^{\Delta (K,m)}$ or, put differently, should have $\Delta (K,m) \in \Z$ rather than $\Delta (K,m) \in \mathbb{Q}$.
While this curious conclusion is natural in the $R$-matrix approach of \cite{Park:2020edg}, it is rather non-obvious in other constructions of $F_K (x,q)$. It would be interesting to explore it further in the framework of 3d-3d correspondence, curve counting, {\it etc.}

This discussion can help us glean what happens when we try to relax the `mod 1' condition in the relation \eqref{dvsDeltamain}. From \eqref{dvsDeltaLens} and \eqref{dvsDelta31int} we know that any generalization of \eqref{dvsDeltamain} should involve explicit dependence on $b \in \text{Spin}^c (M_3)$. The present discussion of surgeries on knots suggests that the best hope is to find a `mod 2' version of \eqref{dvsDeltamain} and further suggests that the $b$-dependence in such a generalization should be linear. Indeed, if the comparison between $\Delta_b (M_3)$ and $d (M_3, b)$ can be reduced to that of a plumbing with the same linking form $Q$, then in comparing \eqref{d-plumbed} with \eqref{Delta-plumbed} we already saw that both are quadratic polynomials in the characteristic vector with the same quadratic term. The linear terms in \eqref{d-plumbed} and \eqref{Delta-plumbed} are only equal mod 1, and therefore would contribute to the relation between $\Delta_b (M_3)$ and $d (M_3, b)$ mod 2. For example, for an integral surgery on the unknot ({\it i.e.} plumbing graph with only one vertex) it shows that a `mod 2' version of \eqref{dvsDeltamain} has an extra term $+b$, consistent with \eqref{dvsDeltaLens}. Generalizing this to arbitrary rational surgeries on other knots requires better control on the overall factors $q^{\Delta (K,m)}$ that we leave to future work.

\section{Outlook and future directions}
\label{sec:conclusions}

The enumerative world of BPS invariants is based on counting geometric objects, solutions to partial differentials equations, and the like.
It is tantalizing to explore new bridges between this world and a different looking world of topological invariants that, on the one hand, has its roots in the pioneering works of Pontryagin, Thom, and others, and, on the other hand, recently found its applications in classifying topological quantum field theories and topological phases of matter.

The particular bridges we found connect BPS $q$-series invariants of 3-manifolds with the Rokhlin invariant and the correction terms of Ozsv\'{a}th and Szab\'{o}. From a broader perspective, it would be interesting to explore similar relations that involve other BPS partition functions and other cobordism invariants, including the invariants of homology cobordisms mentioned in the Introduction. Even in the narrower context of the $q$-series invariants \eqref{Zhatqseries} studied here, quite a few open questions still remain. For example, one obvious question is to test the relations we found, \eqref{dvsDeltamain} and \eqref{muvsZmain}, in more general examples and, if they still hold, to understand why this is the case. To answer this general question it may help to break it into more concrete problems:
\begin{itemize}

\item Combining the tools of \cite{Gukov:2019mnk} and \cite{Park:2020edg}, one can compute the explicit form of the $q$-series \eqref{Zhatqseries} for many hyperbolic surgeries. It would be interesting to test the relations \eqref{dvsDeltamain} and \eqref{muvsZmain} in those examples, either numerically, as in section~\ref{sec:hyperbolic}, or through better understanding of modular properties, following \cite{Cheng:2018vpl,MR4015713,Bringmann:2018ddv,MR4055703}.

\item In studying the relations to correction terms and Rokhlin invariants, we mainly focused on rational homology spheres, {\it i.e.} 3-manifolds with $b_1 (M_3) = 0$. A natural question, therefore, is to explore these relations more fully for 3-manifolds with $b_1 > 0$.

\item Similarly, it is natural to extend our analysis to higher-rank BPS $q$-series invariants $\hat Z^{SU(N)}_b (M_3,q)$ and see if other cobordism invariants of $M_3$ can arise in a way analogous to the correction terms and Rokhlin invariants studied here.

\item Another natural research direction is to use various interrelations between topological invariants in order to understand better relations \eqref{dvsDeltamain} and \eqref{muvsZmain}, or their variants. For example, according to \cite{MR1957829}, $d (M_3,b) = \rho (M_3,b)$ mod 2. Therefore, many results of this paper, including \eqref{dvsDeltamain} can be expressed using the invariant $\rho (M_3,b)$ in place of $d (M_3,b)$. It would be interesting to study whether this alternative formulation can help to relax the `mod 1' condition or shed light on the topological nature of the invariant $\Delta_b (M_3)$, {\it e.g.} ``Is $\Delta_b (M_3)$ itself an invariant of homology cobordisms?''

\item ``3d Modularity'' conjecture states that $\hat Z (M_3,q)$ can be identified with characters of (logarithmic) vertex algebras \cite{Cheng:2018vpl}. In this interpretation, $\Delta_b (M_3)$ is identified with the scaling dimension of the corresponding module. It would be interesting to use this interpretation to learn more about the relation \eqref{dvsDeltamain} as well as $\Delta_b (M_3)$ itself.

\item If the relations \eqref{dvsDeltamain} and \eqref{muvsZmain} continue to hold in other examples, it would be desirable to explore their origins a little deeper. In this respect, since the signature of 4-manifolds that enters the definition of the Rokhlin invariant \eqref{Rokhlindef} often appears in four-dimensional gauge theory, one way to tackle this question could be by realizing Rokhlin invariant of $M_3$ in Kapustin-Witten gauge theory, possibly on $M_3 \times \R_+$ with coupling $\Psi = - \frac{1}{4}$ and Nahm pole boundary conditions \cite{MR2852941}.

\item Once the relations \eqref{dvsDeltamain} and \eqref{muvsZmain} --- or their analogues for other BPS partition functions --- are well tested and well understood, the next natural step is to ask whether they can aid in computing the BPS partition functions.
At the very least, this can help writing the leading term $\hat Z_b (M_3,q) = a_0^{(b)} q^{\frac{1}{2} - d (M_3,b)} + \ldots$ when the calculation of the $q$-series is challenging.
A more ambitious hope is that the surgery formulae that we saw {\it e.g.} in the case of correction terms and Rokhlin invariants at special values of $q$ can provide new insights into cutting-and-gluing relations for the entire $q$-series. We hope this pursuit can eventually lead to much needed surgery techniques for computing Vafa-Witten invariants of 4-manifolds.

\end{itemize}

Throughout the paper, we mostly studied \eqref{dvsDeltamain} and \eqref{muvsZmain} in parallel. However, the BPS $q$-series provides a ``bridge'' between these two relations, as illustrated in Figure~\ref{fig:qplane}. Presumably, it can be viewed as a generalization of the known relations between the Rokhlin invariants and correction terms of the same 3-manifold \cite{MR2425720,MR2503523,MR3802260}.
In particular, for negative definite plumbed 3-manifolds, whose BPS $q$-series invariants were obtained in \cite{Gukov:2017kmk}, in each Spin structure there is a relation between the Neumann-Siebenmann invariant and the corresponding correction term \cite{MR2425720}:
\be
\bar \mu (M_3,s) \; = \; - 4 d (M_3,s).
\ee
Using \eqref{muvsZmain}, combined with \eqref{NSvsRokhlin}, we learn that
\be
\frac{1}{2\pi} \arg \hat Z (M_3)
\Big|_{q=i}
\; = \; \frac{1}{4} + \frac{3}{4} d (M_3)
\qquad \text{mod}~1
\label{Zphased}
\ee
where, for simplicity, we assumed $M_3$ to be a homology sphere. The overall factor $q^{\Delta (M_3)}$ contributes to the left-hand side $\frac{1}{4} \Delta (M_3)$.
This relation is consistent with \eqref{dvsDeltamain} and, in fact, predicts how much the rest of the $q$-series should contribute to the phase on the left-hand side of \eqref{Zphased}. In particular, it tells us that this contribution can not be trivial.

\vskip 1.5cm \noindent
{\it Acknowledgements:} We are especially grateful to Francesca Ferrari, Ciprian Manolescu, and Yi Ni for their help and insightful comments. It is also a pleasure to thank Rob Kirby and Paul Melvin for stimulating discussions and inspiration at the triple-header birthday conference \emph{{``Topology in Dimensions 3, 3.5 and 4''}} in Berkeley (June, 2018).
We also thank Cumrun Vafa for encouragement and comments.
The work of S.G. is supported by the U.S. Department of Energy, Office of Science, Office of High Energy Physics, under Award No. DE-SC0011632, and by the National Science Foundation under Grant No. NSF DMS 1664227.
The~research of S.P. is supported by Kwanjeong Educational Foundation.

\end{document}